\documentclass[preprint2,longabstract]{aastex}  

\usepackage{color}
\usepackage{multirow}

\newcommand{\Swift}{{\it Swift}}
\newcommand{\Chandra}{{\it Chandra}}
\newcommand{\cldy}{Cloudy}
\newcommand{\btxt}[1]{{#1}}

\shorttitle{Pan-chromatic observations of nova LMC 2012}
\shortauthors{Schwarz et al.}

\begin{document}

\title{Pan-chromatic observations of the remarkable nova LMC 2012
\footnote{Based on
observations with the NASA/ESA Hubble Space Telescope obtained at the Space
Telescope Science Institute, which is operated by the Association of
Universities for Research in Astronomy, Incorporated, under NASA contract
NAS5-26555.}}

\author{Greg J. Schwarz}
\affil{American Astronomical Society, 2000 Florida Ave., NW, Suite 300, 
DC 20009-1231, Greg.Schwarz@aas.org}

\author{Steven N. Shore}
\affil{Dipartimento di Fisica "Enrico Fermi", Universita di Pisa, and 
INFN- Sezione Pisa, largo B. Pontecorvo 3, 56127, Pisa, Italy}

\author{Kim L. Page, Julian P. Osborne, Andrew P. Beardmore}
\affil{X-Ray and Observational Astronomy Group, Department of Physics 
and Astronomy, University of Leicester, Leicester, UK, LE1 7RH}

\author{Frederick M. Walter}
\affil{Department of Physics and Astronomy, Stony Brook University,
Stony Brook, NY, 11794-3800} 

\author{Michael F. Bode}
\affil{Astrophysics Research Institute, IC2 Liverpool Science Park, 146 Brownlow Hill, L3 5RF, UK}

\author{Jeremy J. Drake}
\affil{Smithsonian Astrophysical Observatory, MS-3, 60 Garden Street, Cambridge, MA 02138, USA}

\author{Jan-Uwe Ness}
\affil{European Space Astronomy Centre, P.O. Box 78, 28691 Villanueva de la Canada, Madrid, Spain}

\author{Sumner Starrfield}
\affil{School of Earth and Space Exploration, Arizona State University, Tempe, AZ 85287, USA}

\author{Daniel R. Van Rossum}
\affil{Department of Astronomy and Astrophysics, University of Chicago, Chicago, IL 60637, USA}

\author{Charles E. Woodward}
\affil{Minnesota Institute for Astrophysics, University of Minnesota, 116 Church St, SE Minneapolis, MN 55455, USA}

\begin{abstract}

We present the results of an intensive  multiwavelength campaign on nova
LMC 2012.  This nova evolved very rapidly in all observed wavelengths.  The
time to fall two magnitudes in the V band was only 2 days.  In X-rays the
super soft phase began 13$\pm$5 days after discovery and ended around day
50 after discovery.  During the super soft phase, the \Swift/XRT and
\Chandra\ spectra were consistent with the underlying white dwarf being
very hot, $\sim$ 1 MK, and luminous, $\sim$ 10$^{38}$ erg s$^{-1}$.  The
UV, optical, and near-IR photometry showed a periodic variation after the
initial and rapid fading had ended.  Timing analysis revealed a consistent
19.24$\pm$0.03 hr period in all UV, optical, and near-IR bands with
amplitudes of $\sim$ 0.3 magnitudes which we associate with the orbital
period of the central binary.  No periods were detected in the
corresponding X-ray data sets.  A moderately high inclination system,
$i$ = 60$\pm$10$^{\arcdeg}$, was inferred from the early optical
emission lines.  The {\it HST}/STIS UV spectra were highly unusual with
only the \ion{N}{5} (1240\AA) line present and superposed on a blue  
continuum.  The lack of emission lines and the observed UV and optical
continua from four epochs can be fit with a low mass ejection event, $\sim$
10$^{-6}$ M$_{\odot}$, from a hot and massive white dwarf near the
Chandrasekhar limit.  The white dwarf, in turn, significantly illuminated
its subgiant companion which provided the bulk of the observed UV/optical
continuum emission at the later dates.  The inferred extreme white dwarf
characteristics and low mass ejection event favor nova LMC 2012 being a
recurrent nova of the U Sco subclass.

\end{abstract}


\section{Introduction}

Nova explosions occur in binary systems when a white dwarf (WD) accretes a
sufficient amount of mass lost from its companion, either from Roche lobe
overflow in short period systems or a wind in long period systems, to
initiate thermonuclear reactions.  Core material in the WD is
mixed into the accreted layers and the envelope is violently ejected when
the pressure at the degenerate WD-accretion interface becomes high enough
to trigger a thermonuclear runaway (TNR).  Novae thus expel a mixture of
accreted gas, material from the underlying WD, and products of
nucleosynthesis from the TNR.  The amount of mass accreted and subsequently
ejected, and the energetics of the outburst, depend on the WD mass and
composition plus the accretion rate.

While there have been many multiwavelength studies of Galactic
\citep{Sch11} and M31 novae \citep{Henze14}, there are rather few for those
in the Magellanic Clouds.  Novae in the clouds have the advantage of coming
from a more homogeneous population than in the Galaxy and are
effectively at the same distance and low extinction.  Since the LMC is
significantly closer than M31, its novae can be followed well into their
nebular phases.  This is currently impossible for M31 novae since they
generally fade below the optical/NIR background after a decline of only a
few magnitudes.

\begin{figure*}
\includegraphics[scale=0.8]{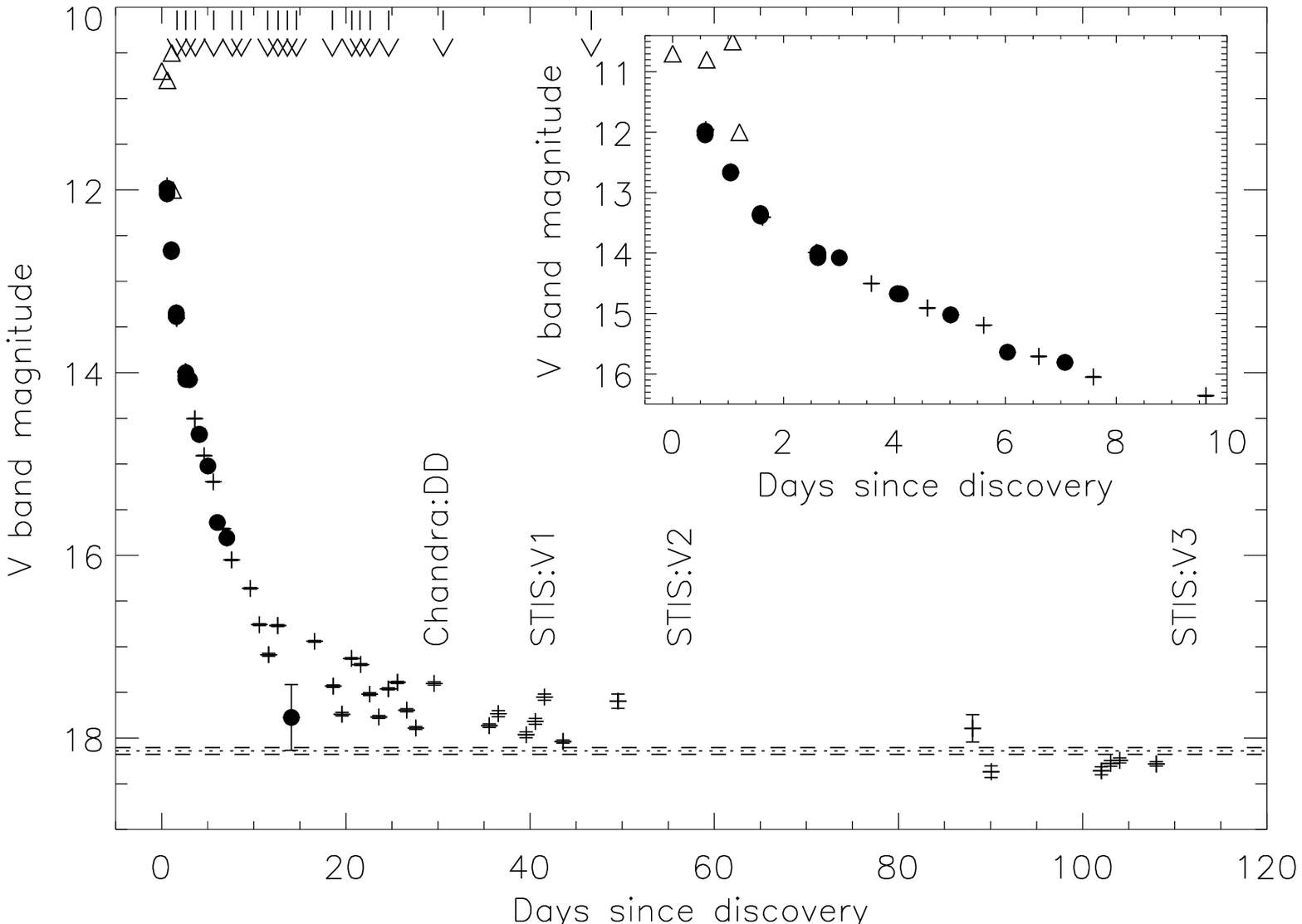}
\caption{The {\it SMARTS} (pluses) and AAVSO (filled circles) V band light curve.
The earliest visual estimates from \citet{CBET3071S} are shown as triangles.
The times of the three {\it HST}/STIS observations, a \Chandra\ DDT observation,
and {\it SMARTS} spectroscopy (downward arrows at top) are also labeled.  The dotted 
line is the V band magnitude of the nearby field star \citep{Z04} with the 
dashed lines showing its uncertainty. The inset shows the early decline in 
greater detail.  \label{Vlc}}
\end{figure*}

Nova LMC 2012 (TCP J04550000-7027150) was discovered on March 26.397 UT
(MJD 56012.897) at a visual magnitude of 10.7 \citep{CBET3071S}.  
The upper limit on the visual magnitude was 12.5 mag twelve days
prior to the Seach discovery.  The discovery date is taken as day zero and
the shorthand ``Dn'', where ``n'' is the number of days after day zero,
is used for all the following observations.  Figure \ref{Vlc} shows the V
band light curve.  The early visual and unfiltered estimates given in
\citet{CBET3071S} are also included for completeness.  The last visual
estimate was about 0.5 magnitudes brighter than the V band observation
taken at the same time.  This is typical of visual magnitudes since they
are more sensitive to H$\alpha$ emission, due to the red sensitivity of the
human eye.  With this correction, we estimate that LMC 2012 likely reached
V maximum on D0.3 at $<$ 11 mag.  The time to decline two and three
magnitudes (t$_{2,3}$) using only the observed V band photometry was 2 and
4.5 days, respectively.  These are upper limits since there was no V band
photometry prior to D0.6 and the decline is faster just after maximum.
These optical declines are similar to V838 Her \citep{Van96} and V4160 Sgr
\citep{Sch07} which makes LMC 2012 one of the photometrically fastest novae
ever observed.

\citet{CBET3071P} obtained an optical spectrum on D0.4.  The spectrum had
strong, broad emission lines with P-Cygni profiles.  The absorption troughs
of the P-Cygni profiles implied a large expansion velocity of $\sim$
5,000 km s$^{-1}$.

After discovery, LMC 2012 was observed by a number of different facilities
at a variety of wavelengths.  This paper reports on the pan-chromatic
observations from \Swift, \Chandra, {\it HST}, and {\it SMARTS}.

\section{Nova position \label{confuse}}

The LMC 2012 position provided by Bohlsen in the AAVSO Special Notice \#270
\footnote{\url{http://www.aavso.org/aavso-special-notice-270}}, was RA
04:54:56.81, Dec -70:26:56.4 (J2000).  This optical position was close to
the position derived from the \Swift/UVOT images when LMC 2012 was bright,
and the {\it HST} derived position.  A Virtual Observatory datascope \footnote{
\url{https://heasarc.gsfc.nasa.gov/cgi-bin/vo/datascope/init.pl}}
positional search within 1 arcsecond of the optical position reveals five
pre-outburst sources.  With the positional uncertainties involved, these
five pre-outburst sources are likely the same object, see Table
\ref{astrometry}.  Four of the five pre-outburst sources have photometry
spanning the ultraviolet to near infrared.  GALEX detected a source with
its NUV detector, $\lambda_{\rm eff}$ = 2267\AA, at 18.94 $\pm$ 0.04 mag.  The
LMC photometric survey catalog \citep{Z04} source has Johnson UBV and Gunn
I photometry of 17.562$\pm$0.048, 18.097$\pm$0.388, 18.140$\pm$0.038, and
18.271$\pm$ 0.054 magnitudes, respectively.  The USNO-B1 catalog source has
R1, B2, and R2 plate magnitudes of 17.76, 21.46, and 18.9 magnitudes,
respectively.  The IRSF Magellanic Cloud point source catalog
\citep{Kato07} has a source with J and H magnitudes of 18.37$\pm$0.07 and
18.62$\pm$0.25 magnitudes, respectively.  The last source is from the Guide
Star Catalog (V2.3) which has red, green and 0.8$\mu$m photographic band
magnitudes of 18.54, 17.69 and 18.52, respectively. The horizontal dotted
and dashed lines in Figure \ref{Vlc} show the \citet{Z04} source V band
magnitude and uncertainty.  By D60, LMC 2012 and the pre-outburst source had
equivalent brightness, which contaminated the nova measurements and masked
its decline at later times.

\begin{figure*}[htbp]
\begin{tabular}{cc}
\includegraphics[scale=0.45]{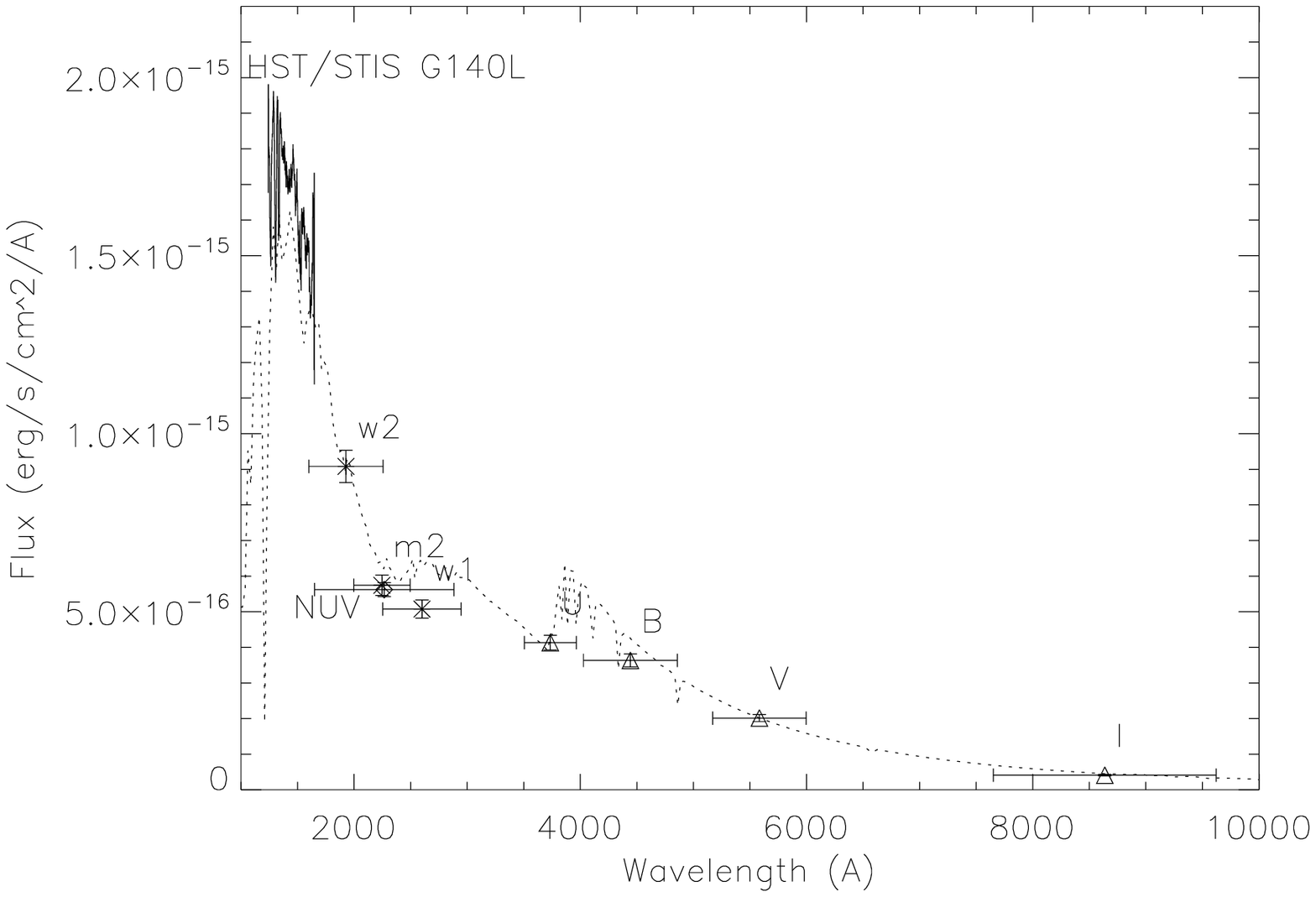} & \includegraphics[scale=0.45]{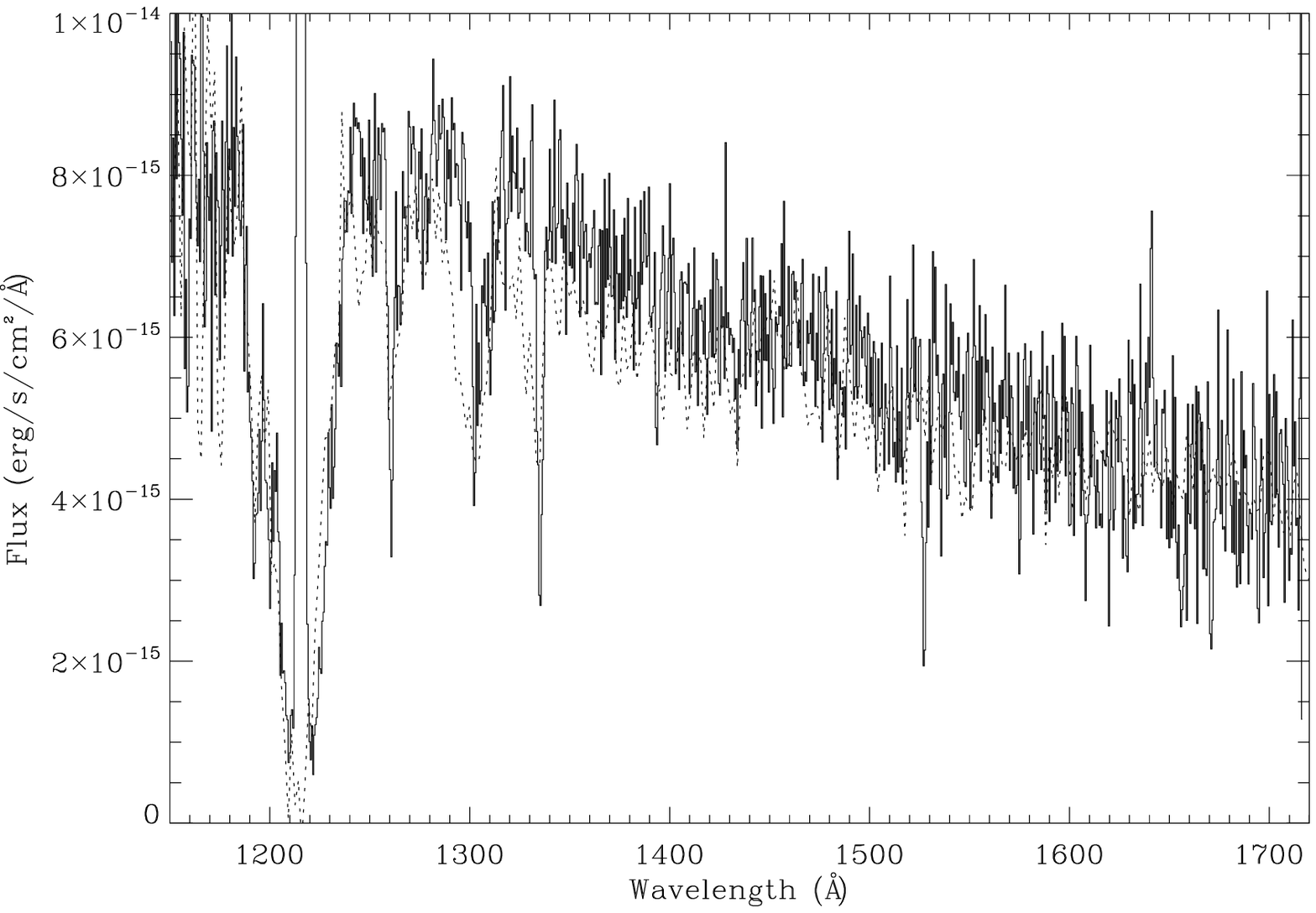} 
\end{tabular}
\caption{Left figure: The \citet{Z04} field star UBVI photometry, the GALEX 
NUV data, quiescent \Swift/UVOT w2, m2, and w1 photometry, and the last 
{\it HST}/STIS G140L spectrum. The best fit is with a 15,000 K, log(g)=4 
\citet{Kurucz} model atmosphere (dotted line).  The model has been dereddened 
with E(B-V) = 0.15 mag \citep{LMCEbmv} and scaled by (r$_{star}$/D)$^2$
where r$_{star}$ = 3.8 R$_{\odot}$ and D = 48 kpc.
Right figure: Comparison of the last {\it HST}/STIS G140L spectrum (solid line;
dereddened by E(B-V) = 0.15) and an IUE spectrum (SWP15537) of the B3 V 
star $\rho$ Aur (dotted line) dereddened by E(B-V) = 0.007 and scaled 
only by the distance difference (162 pc/48 kpc)$^2$.
\label{bstarcomp}}
\end{figure*}

The pre-outburst photometry was supplemented with the \Swift/UVOT
photometry obtained on D303 when the outburst was clearly over (see Section
\ref{swiftmodel}).  By the last {\it HST} visit on D112.3, LMC 2012 was so
faint that the initial acquisition locked on the brighter pre-outburst
source.  The subsequent FUV spectrum is included with the pre- and
post-outburst photometry in Figure \ref{bstarcomp}a.  Assuming a LMC
distance modulus of 18.5 \citep{HSTKey} and an average E(B-V) = 0.15 mag
\citep{LMCEbmv} the available data are consistent with a B5 V at an
effective temperature of 15,000 K. Figure \ref{bstarcomp}b shows that the
FUV spectrum of the acquired target is a very close match with the B3 V
star $\rho$ Aur.

The angular separation between LMC 2012 and this field star is of order
0.23''.  The field star is not likely associated with the LMC 2012
system as an offset of 0.2 arcseconds at the LMC distance corresponds to a
minimum separation of $\sim$ 10$^4$ AU assuming both are coplanar.  This
is far too great for Roche lobe mass transfer and \citet{Krticka14} finds that
B stars with effective temperatures of 15,000 K do not have any significant
line driven mass loss that could effectively transfer material from the
field star to the WD.

\begin{deluxetable}{lllcc}
\tablecaption{Nova and field star astrometry\label{astrometry}}
\tabletypesize{\scriptsize}
\tablewidth{0pt}
\tablehead{\colhead{Name} & \colhead{RA (J2000)} &
\colhead{Dec (J2000)} & \colhead{Offset\tablenotemark{a}} &
\colhead{Offset\tablenotemark{b}}\\
\colhead{} & \colhead{(hh:mm:ss.ss)} & \colhead{(dd:mm:ss.s)} &
\colhead{($\arcsec$)} & \colhead{($\arcsec$)}
}
\startdata
{\it HST}/STIS (this work) & 04:54:56.94 & -70:26:56.43 & \nodata & 0.23 \\
Optical \citep{CBET3071S} & 04:54:56.81 & -70:26:56.4 & 0.05 & 0.20 \\
UVOT (Bright)\tablenotemark{c} & 04:54:56.93 & -70:26:56.1 & 0.33 & 0.10 \\
\hline
UVOT (Faint)\tablenotemark{c} & 04:54:56.84 & -70:26:56.0 & 0.43 & 0.20 \\
UVOT (Quiescence)\tablenotemark{c} & 04:54:56.81 & -70:26:55.96 & 0.47 & 0.24 \\
MC Photometric Survey\tablenotemark{d} & 04:54:56.84 & -70:26:56.1 & 0.33 & 0.10 \\
USNO-B1 0195-0050667\tablenotemark{e} & 04:54:56.89 & -70:26:56.20 & 0.23 & \nodata \\
S1HN007287\tablenotemark{f} & 04:54:56.77 & -70:26:56.24 & 0.20 & 0.06 \\
MCPSC 04545681-7026563\tablenotemark{g} & 04:54:56.82 & -70:26:56.3 & 0.14 & 0.10 \\
GALEX J045456.8-702656\tablenotemark{h} & 04:54:56.82 & -70:26:56.90 & 0.47 & 0.7 \\
\enddata
\tablecomments{Positions above the line are associated with LMC 2012
while positions below the line are for the nearby field star.}
\tablenotetext{a}{Offset from {\it HST}/STIS derived LMC 2012 position.}
\tablenotetext{b}{Offset from USNO-B1 survey field star position.}
\tablenotetext{c}{From sequences 32326006 (MJD = 56030.614; D18.2) and 49549001 (MJD = 56315.621; D303), respectively. UVOT systematic 1$\sigma$ uncertainty is 0.26$\arcsec$.}
\tablenotetext{d}{\citet{Z04}.  Uncertainty $\sim$ 0.45$\arcsec$.}
\tablenotetext{e}{USNO-B1 survey. Uncertainty is $\sigma_{RA}$ = 0.086$\arcsec$ and $\sigma_{Dec}$ = 0.114$\arcsec$.}
\tablenotetext{f}{Guide Star Catalog 2.3. Uncertainty is $\sigma_{RA}$ = 0.286$\arcsec$ and $\sigma_{Dec}$ = 0.251$\arcsec$.}
\tablenotetext{g}{IRSF Magellanic Clouds Point Source Catalog \citep{Kato07}. Uncertainty is approximately 0.1$\arcsec$.}
\tablenotetext{h}{From Nearby Galaxy Survey. Uncertainty is approximately 0.5$\arcsec$ \citep{Mo07}.}
\end{deluxetable}

\section{The pan-chromatic data set}

\subsection{\Swift\ UV and X-ray data}

\Swift\ is a revolutionary facility for studying novae \citep[see][for
details]{Sch11}.  Its three instruments cover the $\gamma$-ray (BAT), X-ray
(XRT), plus UV and optical (UVOT) bandpasses.  The XRT has superb soft X-ray
response \citep{2005SSRv..120..165B} which makes it ideal for observing the
Super-Soft-Source (SSS) phase of novae.  The UVOT provides coincident
UV/optical six filter photometry or low resolution grism
spectroscopy \citep{2005SSRv..120...95R}.  LMC 2012 was not detected by the
BAT but an extensive data set exists from the numerous UVOT and XRT
detections which are described below.

\begin{deluxetable}{ccccc}
\tablecaption{UV, Optical, and IR Photometry\label{phot}}
\tablecolumns{5}
\tabletypesize{\scriptsize}
\tablewidth{0pt}
\tablehead{
\colhead{$\Delta{t}$\tablenotemark{a}} & \colhead{MJD} &
\colhead{mag} & \colhead{$\sigma$} & \colhead{Source} \\
\colhead{(d)} & \colhead{(d)} & \colhead{(mag)} &
\colhead{(mag)} & \colhead{}
}
\startdata
\cutinhead{m2 band}
  1.292 & 56013.689 & 11.297 &  0.051 & \Swift/UVOT \\
  1.338 & 56013.735 & 11.309 &  0.051 & \Swift/UVOT \\
\cutinhead{B band}
  0.597 & 56012.994 & 11.900 &  0.002 & {\it SMARTS} \\
  1.040 & 56013.437 & 12.569 &  0.048 & AAVSO \\
\cutinhead{V band}
  0.584 & 56012.981 & 11.985 &  0.007 & AAVSO \\
  1.044 & 56013.441 & 12.674 &  0.009 & AAVSO \\
\cutinhead{R band}
  0.598 & 56012.995 & 11.278 &  0.002 & {\it SMARTS} \\
  1.047 & 56013.444 & 11.841 &  0.007 & AAVSO \\
\cutinhead{I band}
  0.596 & 56012.993 & 10.932 &  0.003 & {\it SMARTS} \\
  1.053 & 56013.450 & 11.616 &  0.023 & AAVSO \\
\cutinhead{J band}
  0.596 & 56012.993 & 10.407 &  0.027 & {\it SMARTS} \\
  1.618 & 56014.015 & 11.867 &  0.051 & {\it SMARTS} \\
\cutinhead{H band}
  0.596 & 56012.993 &  9.883 &  0.019 & {\it SMARTS} \\
  1.618 & 56014.015 & 11.566 &  0.039 & {\it SMARTS} \\
\cutinhead{K band}
  0.596 & 56012.993 &  9.369 &  0.024 & {\it SMARTS} \\
  1.618 & 56014.015 & 10.827 &  0.048 & {\it SMARTS} \\
\enddata
\tablenotetext{a}{Where t$_0$ is the discovery date, 2012 March 26.397 UT (MJD 56012.897)}
\tablecomments{Table \ref{phot} is published in its entirety in the electronic
edition of the {\it Astronomical Journal}.  A portion is shown here
for guidance regarding its form and content.}
\end{deluxetable}

\subsubsection{\Swift\ UVOT \label{uvot}}

\begin{figure*}
\includegraphics[scale=0.9]{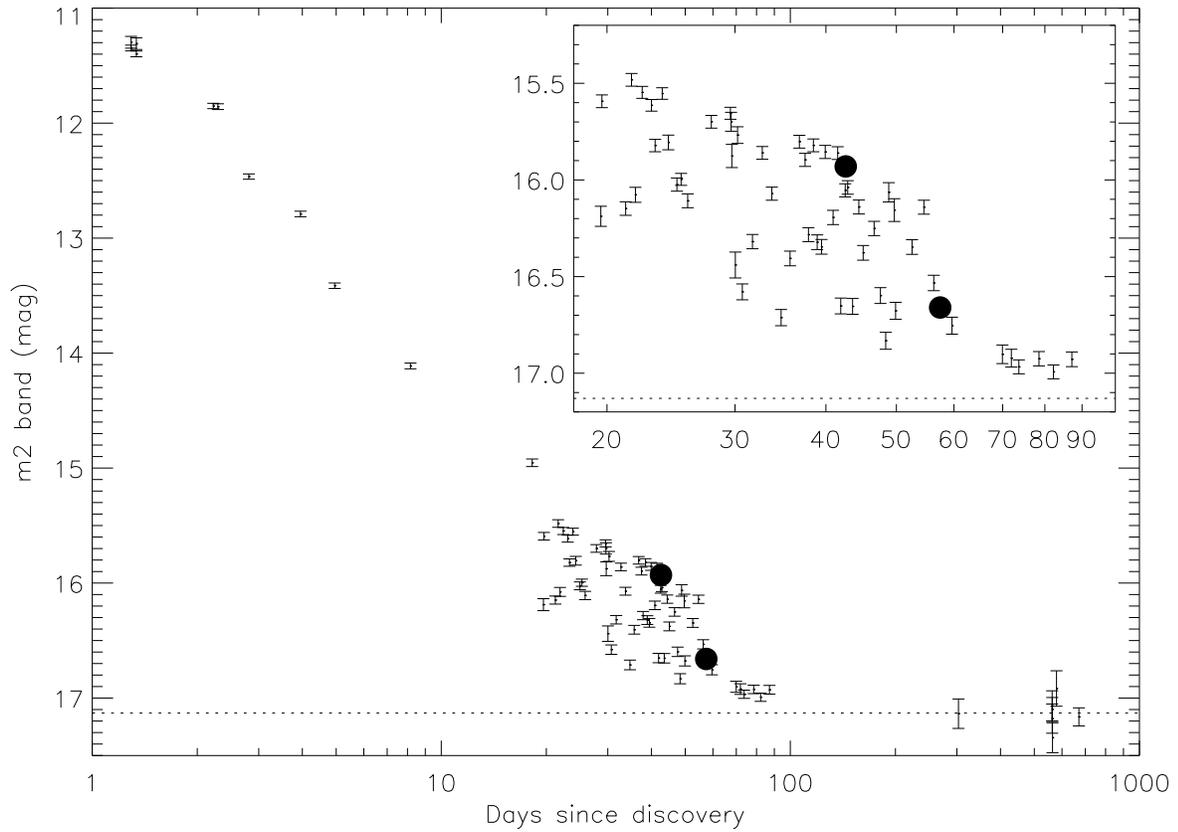}
\caption{The \Swift/UVOT uvm2 band ($\lambda_{\rm eff}$ = 2246\AA) light curve
of LMC 2012.  The filled circles are the derived uvm2 band magnitudes from the
{\it HST} spectra.  The dotted line is the mean magnitude from the post D300
photometry derived from the field star. The inset expands on the significant
short-term variability in the light curve between D20 and D60.
%% HST uvm2 magnitudes are 15.93 and 16.66 mags.
%% Average field star uvm2 magnitude is 17.13 mag.
%% m = Z_pt_ - 2.5*log(CRate) where Z_pt_ = 16.82 for uvm2.
%% Therefore the base count rate, i.e. the field star is 0.75 ct/s
%% The count rate on the 2nd HST visit was 1.16 ct/s. Field star was ~2/3
%% The count rate on the 1st HST visit was 2.27 ct/s.  Field star was ~1/3
\label{swiftm2lc}}
\end{figure*}

\Swift\ obtained 74 uvm2 band ($\lambda_{\rm eff}$ = 2246\AA, FWHM =
498\AA) observations of LMC 2012 with the UVOT instrument from D1.2 until
D671.  There were also 12 uvw2 band ($\lambda_{\rm eff}$ = 1928\AA, FWHM =
657\AA) and 9 uvw1 band ($\lambda_{\rm eff}$ = 2600\AA, FWHM = 693\AA)
observations which were only obtained early on D1.2 and later during the
observations after D300.  The UVOT photometry is provided in Table
\ref{phot} while the uvm2 light curve is shown in Figure \ref{swiftm2lc}.

\Swift\ observed LMC 2012 on five of the first eight days after
discovery. These initial observations revealed that the uvm2 light curve
declined 0.4 mag d$^{-1}$.  For the next ten days, the nova could not be
observed by \Swift\ due to pointing constraints.  When observations resumed
LMC 2012 was a variable UV source with an $\sim$ 0.3 amplitude modulation
superimposed on a declining light curve. It faded from uvm2=15 to uvm2=17
magnitudes between D20 and D60.  The amplitude of the oscillations was
roughly constant from D20 through D50.  From D50 to D60 the amplitude
appears to have decreased to $\sim$ 0.15 mag but there were only 4
observations during this time.  No further significant decline was seen
until D90.  When the same field was subsequently observed after D303 the
detected source was approximately 0.2 magnitudes fainter and within the
uncertainty remained constant at 17.13 mag. The averaged uvm2 magnitude is
shown as a dotted line in Figure \ref{swiftm2lc} and represents the UV
contribution of the field star.  Figure \ref{swiftm2lc} suggests that LMC
2012 dominated the field star prior to D100.

\subsubsection{XRT \label{xrtsection}}

\Swift\ observations of LMC 2012 began on D1.3 and continued for over
670 days.  The XRT spectra were extracted for each individual observation
using the latest version of the \Swift\ software (HEASOFT 6.15.1 and the v014
calibration file for the PC RMF).  These had typical exposure times of
$\sim$ 1 ks, with one or two observations occurring most days between D18
and D54, and then once every 2-5 days until D87. All the data were
collected using Photon Counting mode and circular regions were used for
both source and background spectral extraction. If the source count rate
was above about 0.6 count s$^{-1}$, the core of the Point Spread Function
was excluded in order to avoid pile-up.  An 8 pixel excluded radius was used
until D50.  After that, as the nova faded, it was reduced to 5 pixels and
then to zero after D55.  Event grades 0-12 were used for the timing
analysis (Section \ref{periodanal}) but only grade 0 were used for the 
spectral fitting (Section \ref{swiftmodel}) to minimize residual pile-up.

\begin{figure*}
\includegraphics[scale=0.9]{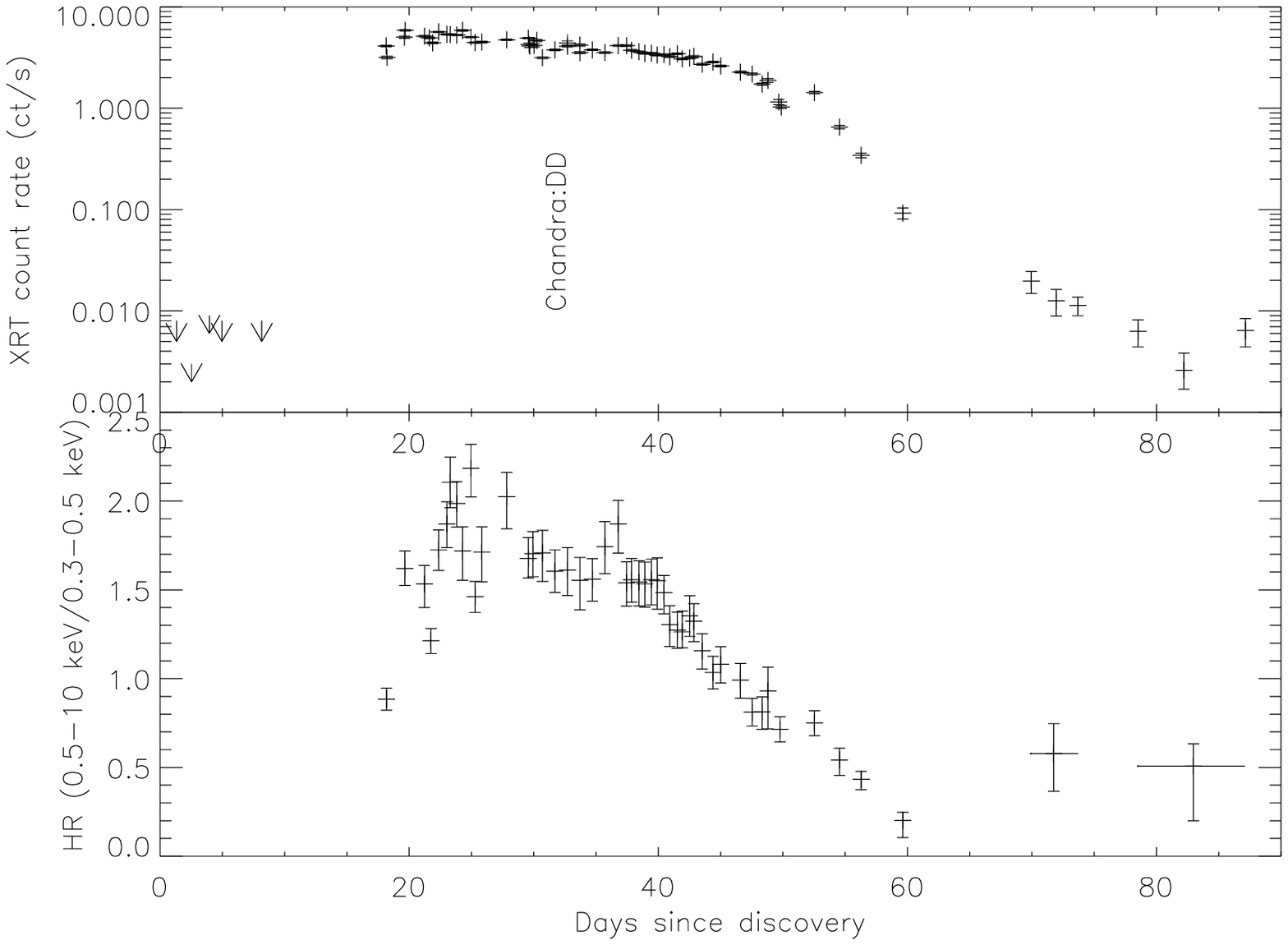}
\caption{The \Swift/XRT 0.3-10 keV count rate (top) and hardness ratio 
([0.5-10 keV/0.3-0.5 keV]; bottom) evolution.  Upper limits (3$\sigma$)
to the count rate (top) are indicated by downward arrows.
The time of the \Chandra\ visit is shown to guide the eye.
\label{swiftXraylc}}
\end{figure*}

Figure \ref{swiftXraylc} shows the 0.3 - 10 keV XRT count rates and
hardness ratios around the epoch when it was X-ray active.  The data are
provided in Table \ref{xrtlc}.  The upper limits (3$\sigma$) and detection
uncertainties (1$\sigma$), when the count rate was $<$0.01 ct s$^{-1}$,
were calculated using Bayesian statistics.

There was no X-ray detection in the first five observations through D8.2
when the UVOT recorded its rapid UV decline.  Once LMC 2012 was no longer
pointing constrained, monitoring began again and the next \Swift\
observation on D18.15 detected a bright and soft X-ray source with a count
rate of 4.12 $\pm$ 0.08 ct s$^{-1}$.

\begin{figure*}
\includegraphics[angle=-90,scale=0.60]{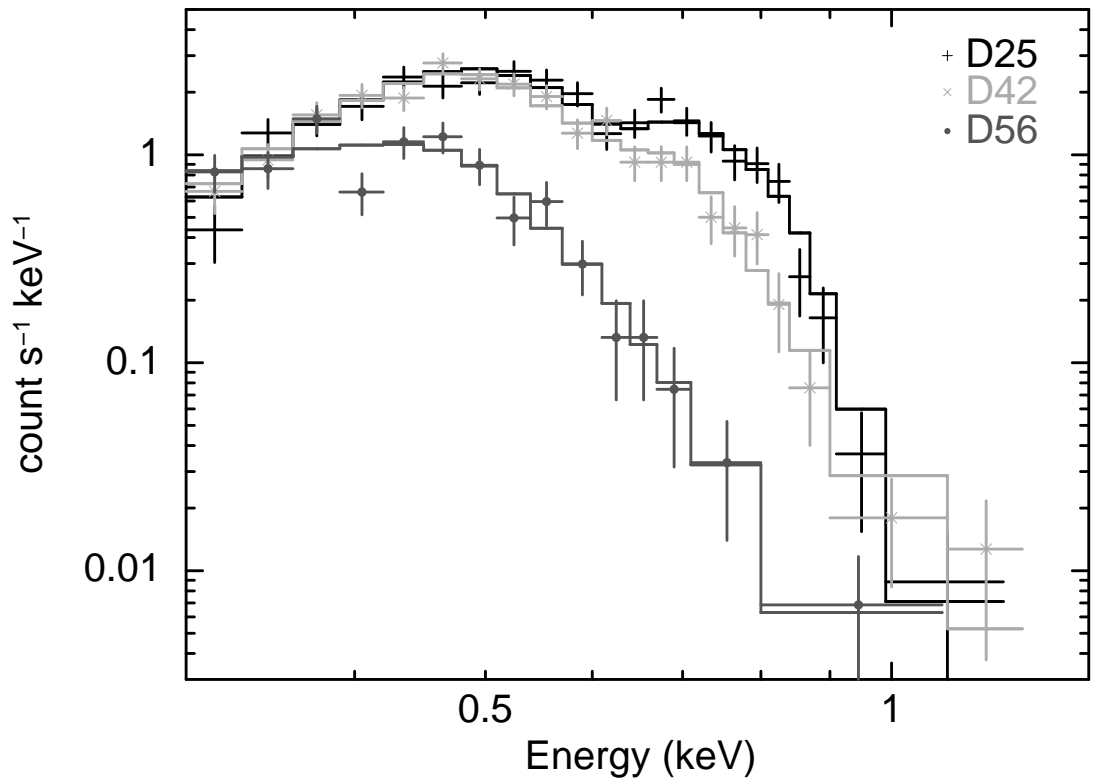}
\caption{\Swift/XRT spectra of LMC 2012 obtained at X-ray maximum (D25; pluses) 
and nearly coincident with the first (D42; Xs) and second (D56; filled circle) 
{\it HST} observations.  The spectra are fit with the models described in Figure
\ref{swiftfit} but also include an optically-thin MEKAL model to account for
the counts above 1 keV. The spectral evolution is consistent with a fading
and cooling source.
\label{xrtspectra}}
\end{figure*}

LMC 2012 reached a maximum X-ray count rate of $\sim$ 6 ct s$^{-1}$ on
D19.7.  The hardness ratio, centered on the soft component, (HR =
[0.5-10 keV]/[0.3-0.5 keV]) at maximum was about 2 at D25. This 
hardness ratio indicates a soft X-ray spectrum consistent with hot
thermal emission at kT $\sim$ 90 eV (see section \ref{swiftmodel}).  The
X-ray source remained approximately constant in both the flux and hardness
ratio over the next 20 days, after which it began declining in both.
Figure \ref{xrtspectra} shows the spectral evolution using the XRT spectra
obtained on D25, D42 and D56.  By D80 the XRT count rate had declined by a
factor of 100 and the nova had become a significantly softer X-ray source
than that seen around D60 as the spectral energy distribution
shifted to cooler temperatures.  Additional \Swift\ observations obtained
starting on D303 and ending on D671 detected no X-ray source.

LMC 2012 entered its SSS phase after D8 and before D18 so we set the SSS
turn-on time, t$_{\rm on}$, to 13$\pm$5 days.  This is a very rapid turn-on
and implies either that very little mass was expelled or the ejecta were
significantly aspherical \citep{Shore13} since the X-ray turn-on is due to
the decrease in the optical depth of the ejecta, for a given ejection
velocity. An upper limit on the  ejected mass of order 10$^{-6}$
M$_{\odot}$ is suggested from the expansion velocity and t$_{\rm on}$
\btxt{based on the simple homogenous and uniformly expanding shell models}
of \citet[][see their Figure 8]{Sch11}.  This low mass is very similar to
those of fast recurrent novae \citep[for example,
see][]{Schaefer11,Anupama13}.

Based on its X-ray light curve and hardness ratio, LMC 2012 ended its SSS
phase around D50.  The SSS duration is inversely proportional to the WD
mass and a short timescale implies a high mass WD \citep{Starr91}.
Compared to the turn-off times, t$_{\rm off}$ of Table 5 in \citet{Sch11},
LMC 2012 had one of the fastest X-ray turn-off times detected.  Figure 9 in
\citet{Sch11} shows how rare this rapid a turn-off is relative to the
Galactic novae with SSS detections.  A t$_{\rm off}$ = 50 days is similar
to recurrent novae such as RS Oph \citep[60 days;][]{Osb11}, V745 Sco
\citep[$\sim$ 5 days;][]{ATel5897}, and U Sco \citep[34 days;][]{ATel2477}
plus the suspected recurrents V2491 Cyg \citep[44 days;][]{Page10}, V2672
Cyg \citep[28 days;][]{Sch11}, and V407 Cyg \citep[30 days;][]{Sch11}.
There are also novae in M31 with similar rapid turn-off times
such as the recurrent M31 2008-12a (t$_{\rm off} \sim$ 19 days)
\citep[see][for details]{Henze14}.

\citet{Henze14} compiled four correlations between X-ray and nova
properties for M31 novae.  That galaxy is ideal for these sorts of
comparisons since the uncertainties in the distance are effectively
eliminated and there are sufficient numbers detected each year to create a
statistically viable sample. \btxt{These observational} relations (their
eqs. 4-7) give the t$_{\rm off}$ vs t$_{\rm on}$, t$_{\rm off}$ vs the
effective blackbody temperature, t$_{\rm on}$ vs t$_{2,R}$, and t$_{\rm
on}$ vs v$_{exp}^{max}$ relationships. For LMC 2012 we adopted t$_{\rm on}$
= 13 days, t$_{\rm off}$ = 50 days, kT = 86 eV from the \Swift/XRT model
fits, v$_{exp}$ = 5,000 km s$^{-1}$ from the estimates from the early
P-Cygni absorption lines, and t$_{2,R} \sim$ 2.5 days from Figure
\ref{Vlc}. The observed X-ray behavior of LMC 2012 is well described by
these equations.  Its predicted t$_{\rm off}$ times from the
\citet{Henze14} equations 4 and 5 are 62$^{+41}_{-16}$ and 71$^{+14}_{-13}$
days, respectively.  The derived t$_{\rm on}$ times for equations 6 and 7
are both $\sim$ 14$^{+5}_{-4}$ days.

\begin{deluxetable}{rcl|rcl}
\tablecaption{\Swift/XRT 0.3 - 10 keV count rate and hardness ratio\label{xrtlc}}
\tabletypesize{\scriptsize}
\tablecolumns{6}
\tablewidth{0pt}
\tablehead{
\colhead{$\Delta{t}$\tablenotemark{a}} & \colhead{MJD} & \colhead{CR} &
\colhead{$\Delta{t}$\tablenotemark{a}} & \colhead{MJD} & 
\colhead{HR\tablenotemark{b}} \\
\colhead{(d)} & \colhead{(d)} & \colhead{(ct/s)} &
\colhead{(d)} & \colhead{(d)} & \colhead{} 
}
\startdata
 1.315 & 56013.715 & $<$0.008                    & & & \\
 2.525 & 56014.922 & $<$0.003                    & & & \\
 3.953 & 56016.352 & $<$0.009                    & & & \\
 4.957 & 56017.355 & $<$0.008                    & & & \\
 8.162 & 56020.559 & $<$0.008                    & & & \\
18.149 & 56030.547 & 4.12$^{+0.08}_{-0.08}$ & \multirow{2}{*}{18.182} & \multirow{2}{*}{56030.582} & \multirow{2}{*}{0.88$^{+0.06}_{-0.06}$}\\
18.217 & 56030.613 & 3.17$^{+0.09}_{-0.09}$ &        &           & \\
19.614 & 56032.012 & 5.03$^{+0.12}_{-0.12}$ & \multirow{2}{*}{19.649} & \multirow{2}{*}{56032.047} & \multirow{2}{*}{1.62$^{+0.10}_{-0.10}$}\\
19.682 & 56032.082 & 5.89$^{+0.09}_{-0.09}$ &        &           & \\
21.220 & 56033.617 & 5.15$^{+0.08}_{-0.08}$ & 21.220 & 56033.617 & 1.533$^{+0.11}_{-0.13}$\\
21.617 & 56034.016 & 4.95$^{+0.09}_{-0.09}$ & \multirow{2}{*}{21.751} & \multirow{2}{*}{56034.148} & \multirow{2}{*}{1.21$^{+0.07}_{-0.07}$}\\
21.885 & 56034.281 & 4.43$^{+0.09}_{-0.09}$ &        &           & \\
22.362 & 56034.762 & 5.67$^{+0.08}_{-0.08}$ & 22.362 & 56034.762 & 1.73$^{+0.11}_{-0.12}$\\
23.028 & 56035.426 & 5.35$^{+0.07}_{-0.07}$ & 23.028 & 56035.426 & 1.87$^{+0.13}_{-0.13}$\\
23.295 & 56035.695 & 5.34$^{+0.07}_{-0.07}$ & 23.295 & 56035.695 & 2.11$^{+0.14}_{-0.14}$\\
23.822 & 56036.219 & 5.28$^{+0.07}_{-0.07}$ & 23.822 & 56036.219 & 1.97$^{+0.12}_{-0.13}$\\
24.287 & 56036.688 & 5.87$^{+0.11}_{-0.11}$ & 24.287 & 56036.688 & 1.72$^{+0.14}_{-0.17}$\\
24.957 & 56037.355 & 5.03$^{+0.07}_{-0.07}$ & 24.957 & 56037.355 & 2.19$^{+0.13}_{-0.16}$\\
25.294 & 56037.691 & 4.45$^{+0.06}_{-0.06}$ & 25.294 & 56037.691 & 1.46$^{+0.09}_{-0.09}$\\
25.826 & 56038.227 & 4.50$^{+0.09}_{-0.09}$ & 25.826 & 56038.227 & 1.71$^{+0.14}_{-0.17}$\\
27.829 & 56040.227 & 4.75$^{+0.08}_{-0.08}$ & 27.829 & 56040.227 & 2.02$^{+0.14}_{-0.18}$\\
29.566 & 56041.965 & 4.92$^{+0.07}_{-0.07}$ & 29.566 & 56041.965 & 1.68$^{+0.12}_{-0.11}$\\
29.627 & 56042.027 & 4.26$^{+0.13}_{-0.13}$ & \multirow{3}{*}{29.935} & \multirow{3}{*}{56042.332} & \multirow{3}{*}{1.70$^{+0.12}_{-0.13}$} \\ 
29.694 & 56042.094 & 4.12$^{+0.15}_{-0.15}$ &        &           & \\
30.029 & 56042.426 & 4.18$^{+0.17}_{-0.17}$ &        &           & \\
30.242 & 56042.641 & 4.67$^{+0.12}_{-0.12}$ &        &           & \\
30.700 & 56043.098 & 3.16$^{+0.06}_{-0.06}$ & 30.700 & 56043.098 & 1.71$^{+0.13}_{-0.16}$ \\
31.702 & 56044.102 & 3.77$^{+0.06}_{-0.06}$ & 31.702 & 56044.102 & 1.61$^{+0.12}_{-0.12}$ \\
32.699 & 56045.098 & 4.41$^{+0.21}_{-0.21}$ &        &           & \\
32.705 & 56045.102 & 4.10$^{+0.07}_{-0.07}$ & 32.705 & 56045.102 & 1.61$^{+0.13}_{-0.14}$\\
33.702 & 56046.102 & 4.20$^{+0.11}_{-0.11}$ &        &           & \\
33.708 & 56046.105 & 3.54$^{+0.07}_{-0.07}$ & 33.708 & 56046.105 & 1.55$^{+0.13}_{-0.17}$\\
34.708 & 56047.105 & 3.77$^{+0.06}_{-0.06}$ & 34.708 & 56047.105 & 1.56$^{+0.12}_{-0.12}$\\
35.709 & 56048.109 & 3.55$^{+0.06}_{-0.06}$ & 35.709 & 56048.109 & 1.74$^{+0.14}_{-0.15}$\\
36.778 & 56049.176 & 4.16$^{+0.07}_{-0.07}$ & 36.778 & 56049.176 & 1.87$^{+0.13}_{-0.16}$\\
37.446 & 56049.844 & 4.15$^{+0.07}_{-0.07}$ & 37.446 & 56049.844 & 1.54$^{+0.12}_{-0.13}$\\
37.846 & 56050.246 & 3.74$^{+0.06}_{-0.06}$ & 37.846 & 56050.246 & 1.56$^{+0.12}_{-0.13}$\\
38.448 & 56050.848 & 3.61$^{+0.06}_{-0.06}$ & 38.448 & 56050.848 & 1.54$^{+0.12}_{-0.13}$\\
38.915 & 56051.312 & 3.48$^{+0.06}_{-0.06}$ & 38.915 & 56051.312 & 1.53$^{+0.12}_{-0.13}$\\
39.451 & 56051.848 & 3.50$^{+0.06}_{-0.06}$ & 39.451 & 56051.848 & 1.56$^{+0.12}_{-0.14}$\\
39.918 & 56052.316 & 3.35$^{+0.07}_{-0.07}$ & 39.918 & 56052.316 & 1.55$^{+0.13}_{-0.16}$\\
40.452 & 56052.852 & 3.38$^{+0.05}_{-0.05}$ & 40.452 & 56052.852 & 1.49$^{+0.10}_{-0.12}$\\
40.918 & 56053.316 & 3.23$^{+0.06}_{-0.06}$ & 40.918 & 56053.316 & 1.30$^{+0.18}_{-0.12}$\\
41.521 & 56053.918 & 3.45$^{+0.06}_{-0.06}$ & 41.521 & 56053.918 & 1.27$^{+0.10}_{-0.10}$\\
41.921 & 56054.320 & 3.07$^{+0.05}_{-0.05}$ & 41.921 & 56054.320 & 1.26$^{+0.12}_{-0.09}$\\
42.522 & 56054.922 & 3.13$^{+0.06}_{-0.06}$ & 42.522 & 56054.922 & 1.35$^{+0.11}_{-0.12}$\\
42.857 & 56055.254 & 3.25$^{+0.06}_{-0.06}$ & 42.857 & 56055.254 & 1.32$^{+0.10}_{-0.12}$\\
43.524 & 56055.922 & 2.72$^{+0.05}_{-0.05}$ & 43.524 & 56055.922 & 1.16$^{+0.10}_{-0.10}$\\
44.394 & 56056.793 & 2.86$^{+0.05}_{-0.05}$ & 44.394 & 56056.793 & 1.04$^{+0.09}_{-0.09}$\\
45.008 & 56057.406 & 2.62$^{+0.05}_{-0.05}$ & 45.008 & 56057.406 & 1.08$^{+0.10}_{-0.11}$\\
46.597 & 56058.996 & 2.27$^{+0.05}_{-0.05}$ & 46.597 & 56058.996 & 0.99$^{+0.09}_{-0.10}$\\
47.532 & 56059.930 & 2.17$^{+0.05}_{-0.05}$ & 47.532 & 56059.930 & 0.81$^{+0.08}_{-0.08}$\\
48.333 & 56060.730 & 1.73$^{+0.04}_{-0.04}$ & 48.333 & 56060.730 & 0.81$^{+0.09}_{-0.10}$\\
48.804 & 56061.203 & 1.88$^{+0.08}_{-0.08}$ & 48.804 & 56061.203 & 0.93$^{+0.13}_{-0.21}$\\
49.670 & 56062.066 & 1.15$^{+0.07}_{-0.07}$ & \multirow{2}{*}{49.774} & \multirow{2}{*}{56062.172} & \multirow{2}{*}{0.71$^{+0.07}_{-0.07}$} \\
49.875 & 56062.273 & 1.03$^{+0.03}_{-0.03}$ &        &           &                          \\
52.542 & 56064.941 & 1.43$^{+0.04}_{-0.04}$ & 52.542 & 56064.941 & 0.75$^{+0.07}_{-0.07}$\\
54.546 & 56066.945 & 0.65$^{+0.03}_{-0.03}$ & 54.546 & 56066.945 & 0.54$^{+0.07}_{-0.09}$\\
56.282 & 56068.680 & 0.34$^{+0.02}_{-0.02}$ & 56.282 & 56068.680 & 0.43$^{+0.05}_{-0.06}$\\
59.626 & 56072.023 & 0.09$^{+0.01}_{-0.01}$ & 59.626 & 56072.023 & 0.20$^{+0.05}_{-0.10}$\\
69.945 & 56082.344 & 0.02$^{+0.01}_{-0.01}$ & \multirow{2}{*}{71.749} & \multirow{2}{*}{56081.648} & \multirow{2}{*}{0.58$^{+0.17}_{-0.21}$} \\
71.953 & 56084.352 & 0.01$^{+0.004}_{-0.004}$ &        &           & \\
73.683 & 56086.082 & 0.01$^{+0.002}_{-0.002}$ &        &           & \\
78.521 & 56090.918 & 0.01$^{+0.002}_{-0.002}$ &        &           & \\
82.205 & 56095.102 & 0.003$^{+0.001}_{-0.001}$ & 82.966 & 56095.363 & 0.51$^{+0.13}_{-0.3}$\\
87.145 & 56099.543 & 0.006$^{+0.002}_{-0.002}$ &        &           & \\
303.224 & 56315.621 & $<$0.03                    & & & \\
562.965 & 56575.364 & $<$0.01                    & & & \\
563.254 & 56575.651 & $<$0.02                    & & & \\
564.305 & 56576.702 & $<$0.01                    & & & \\
579.227 & 56591.626 & $<$0.04                    & & & \\
671.512 & 56683.910 & $<$0.01                    & & & \\
\enddata
\tablenotetext{a}{Where t$_0$ is the discovery date, 2012 March 26.397 UT (MJD 56012.897)}
\tablenotetext{b}{Where HR = (0.5-10 keV)/(0.3-0.5 keV).}
\tablecomments{On some dates the HR was determined by summing the source
counts from multiple exposures.}
\end{deluxetable}

\subsection{Chandra X-ray spectroscopy \label{chandrasection}}

A Director's Discretionary Time observation of LMC 2012 was obtained with
the Low Energy Transmission Grating (LETG) and High Resolution Camera
Spectroscopic detector (HRC-S).  Observation ID
\dataset[ADS/Sa.CXO#20190]{14426} commenced at UT April 26, 21:56 (D32.016)
and ended at UT 04:00 on 2012 April 27 (D32.270), and had a net exposure
time of 20~ks.  Data were obtained from the {\it Chandra}
archive\footnote{http://cxc.harvard.edu/cda/} and were reprocessed using
CIAO and calibration database versions 4.6.1.  Effective areas and
instrument response files were generated using standard CIAO procedures.

\begin{figure*}
\includegraphics[scale=0.80]{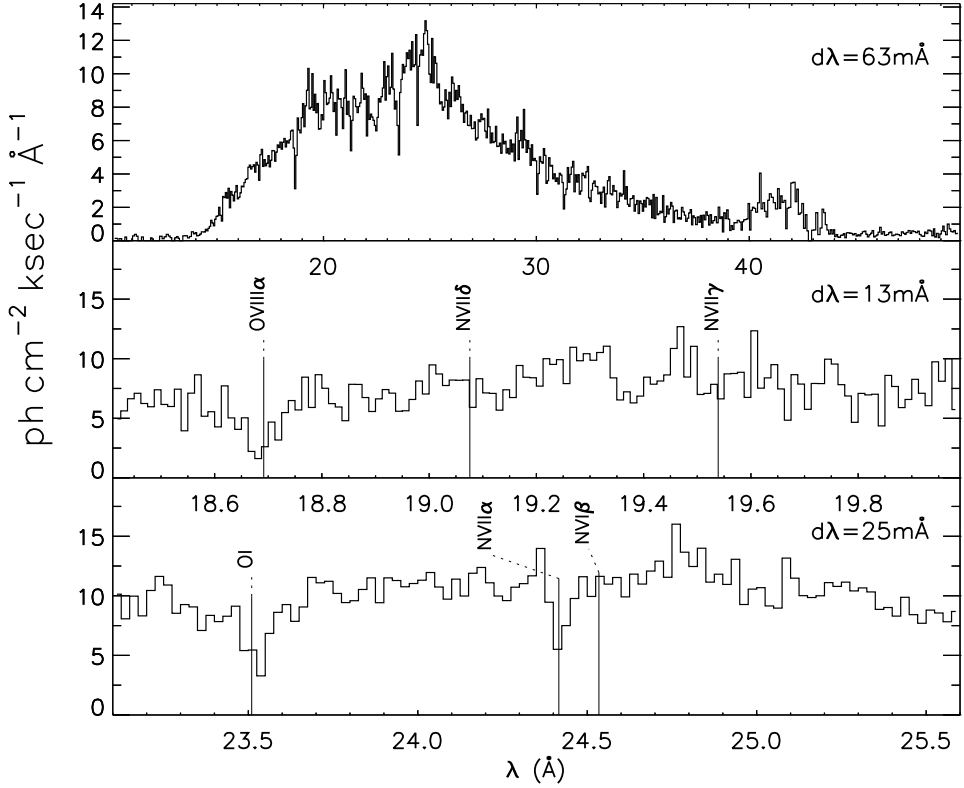}
\caption{Full \Chandra\ DDT spectrum (D32) rebinned to 0.063 \AA\ 
(top panel).  The middle and bottom panels show the absorption and 
emission features around 19 and 24 \AA, respectively.  Various lines 
are labeled and have been Doppler shifted by -4,400 km s$^{-1}$.  
The middle panel spectrum is rebinned to 0.013 \AA\ while 
the bottom panel spectrum is rebinned to 0.025 \AA.
\label{chandramodel}}
\end{figure*}

The combined plus and minus order spectra are shown in Figure
\ref{chandramodel}. Initial reports were provided by \citet{ATel4116} and
\citet{Orio12}.  The spectrum was that of a soft X-ray source with some
emission and absorption lines.  It was exceptionally hot and similar to
high resolution spectra of RS Oph in the supersoft phase (Ness et al 2007),
which had an estimated effective temperature of about $10^6$~K
\citep{Osb11}.  The absorption lines were weaker in LMC 2012 than in the
RS~Oph spectra.  This may be due to the lower metallicity of the LMC,  but
this hypothesis requires detailed model atmosphere analysis to confirm.

The strongest emission lines are the $n=2 \rightarrow 1$
(Lyman-$\alpha$-like) transitions of the hydrogenic ions \ion{N}{7}
$\lambda 24.78$ and \ion{O}{8} $\lambda 18.97$. No prominent features due
to carbon were seen. The \ion{N}{7} and \ion{O}{8} lines exhibited 
P-Cygni-like absorption, blue-shifted by approximately 4400 km~s$^{-1}$.
While this absorption shift is consistent with earlier optical
spectroscopic data, these features could also be a chance superposition of
absorption and emission lines.  The spectral regions around the
\ion{O}{8}$\alpha$ line and \ion{N}{7} lines are illustrated in the middle
and bottom portions of Figure \ref{chandramodel}.  

\begin{figure*}
\includegraphics[scale=0.70]{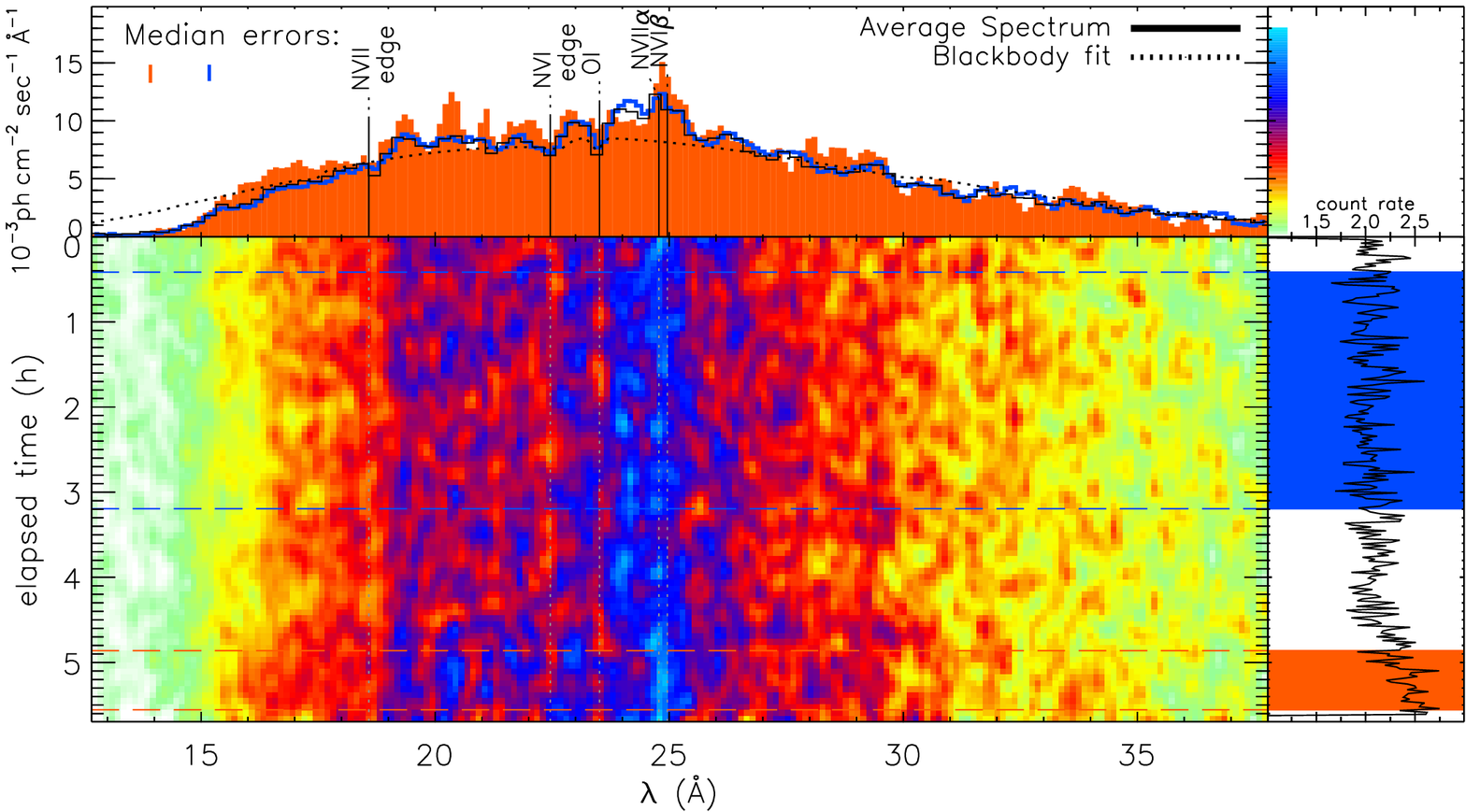}
\caption{Spectral time map of the \Chandra\ observation. The X-ray light 
curve, extracted from the zeroth order, is shown in the right panel,
rotated by 90 degrees. Along the same, downward, time axis, the spectral
evolution is illustrated as a brightness color map in the central panel in
which wavelength runs across (for meaning of color scheme, see top right).
Along the same wavelength axis, selected spectra are shown in the top
panel, with the red and blue colors corresponding to the shaded areas in
the right panel and dashed horizontal lines in the central panel, which
mark the time intervals from which the spectra were extracted. The median
error bars for each spectra are given with the red and blue vertical lines
in the top left corner.  In addition to the two individual spectra, the
average spectrum and a blackbody fit are shown with black and solid and
dotted lines, respectively. \btxt{An increase in count rate by $\sim 25$\%
coincides with higher emission shortward of the NVII absorption edge at
18.6\AA, indicating that reduced absorption causes the higher count rate
(see text). \label{smap}}}
\end{figure*}

In Figure \ref{smap} we show the \btxt{total count and spectral} evolution
of the \Chandra/LETG observation. The observation was divided into 41
adjacent time intervals, each of 500 s duration, from which time-filtered
spectra were extracted in photon flux units. These spectra are arranged in
the central time map with time running down, wavelength across, and
brightness encoded with the color scheme outlined in the top right corner.
The light curve shows a slow \btxt{and small} increase starting at $\sim 4$
hours after the start of the observation.  This increase seems to be
accompanied by \btxt{a limited} increase in the Wien tail, shortward of $\sim
19$\,\AA\ which is close to the N\,{\sc vii} ionization edge at 18.6\,\AA.
The spectrum extracted between 0.4 and 3.2 hours into the observation
(blue) contains a slightly deeper absorption feature at this
wavelength, and the higher flux shortward of 19\,\AA\ in the spectrum
extracted later between 4.9 and 5.5 hours (red) might be due to reduced
absorption caused by N\,{\sc vii}.

Exploration of the full range of model atmosphere parameters to provide
detailed estimates of the element abundances and mass loss rate of LMC~2012
requires extensive and detailed computations that are beyond the scope of
the present work.  \btxt{The goal here is instead to use the {\it Chandra}/LETG
spectrum to support the \Swift\ dataset, obtain an approximate description 
of the global spectral energy distribution, and characterize the ionizing 
flux shortward of the Lyman edge for the photoionization modeling.}

\subsection{HST/STIS spectroscopy \label{hstdata}}

\begin{deluxetable}{cccrrcccl}
\tablecaption{{\it HST} observation log\label{hstlog}}
\tabletypesize{\scriptsize}
\tablecolumns{9}
\tablewidth{0pt}
\tablehead{
\colhead{Exp. ID} & \colhead{UT start time} & \colhead{MJD start time} &
\colhead{$\Delta{t}$\tablenotemark{a}} &
\colhead{Total exp.} & \colhead{Grating} & \colhead{Aperture} &
\colhead{Range} & \colhead{Int. Flux} \\
\colhead{} & \colhead{(hh:mm:ss)} & \colhead{(d)} & \colhead{(s)} &
\colhead{(d)} &
\colhead{} & \colhead{($\arcsec$)} & \colhead{(\AA)} & 
\colhead{erg cm$^{-2}$ s$^{-1}$}
}
\startdata
\cutinhead{Visit 1: 2012-05-07|08}
obtg01010 & 23:30:15 & 56055.479 & 42.58 & 724 & E140M & 0.2X0.2 & 1140 - 1735 & 1.7$\times$10$^{-12}$ \\
obtg01020 & 23:48:24 & 56055.492 & 42.59 & 724 & E230M & 0.2X0.2 & 1574 - 2382 & 9.5$\times$10$^{-14}\tablenotemark{b}$ \\
obtg01030 & 00:05:53 & 56055.504 & 42.61 & 724 & E230M & 0.2X0.2 & 2303 - 3133 & 9.9$\times$10$^{-13}\tablenotemark{c}$ \\
\cutinhead{Visit 2: 2012-05-23}
obtg02010 & 07:30:18 & 56070.313 & 57.41 & 690 & G140L & 52X0.2 & 1140 - 1735 & 1.0$\times$10$^{-12}$ \\
obtg02020 & 07:47:55 & 56070.325 & 57.43 & 690 & G230L & 52X0.2 & 1570 - 3180 & 7.9$\times$10$^{-13}\tablenotemark{d}$ \\
obtg02030 & 08:03:01 & 56070.335 & 57.44 & 780 & G430L & 52X0.2E1 & 2900 - 5700 & 1.0$\times$10$^{-12}$ \\
\cutinhead{Visit 3: 2012-07-17}
obtg99010 & 04:13:43 & 56125.176 & 112.27 & 2746 & G140L & 0.2X0.2 & 1140 - 1735& 9.2$\times$10$^{-13}$  \\
\enddata
\tablenotetext{a}{Where t$_0$ is the discovery date, 2012 March 26.397 UT (MJD 56012.897)}
\tablenotetext{b}{From 1735-2350 \AA.}
\tablenotetext{c}{From 2350-3130 \AA.}
\tablenotetext{d}{From 1735-3130 \AA.}
\end{deluxetable}

After discovery, LMC 2012 was selected as the ToO target of a cycle 19
program (GO-12484) to obtain high resolution UV spectroscopy at
three separate times during its evolution.  The {\it HST} observation log of the
observations is presented in Table \ref{hstlog}.

\begin{figure}
\plotone{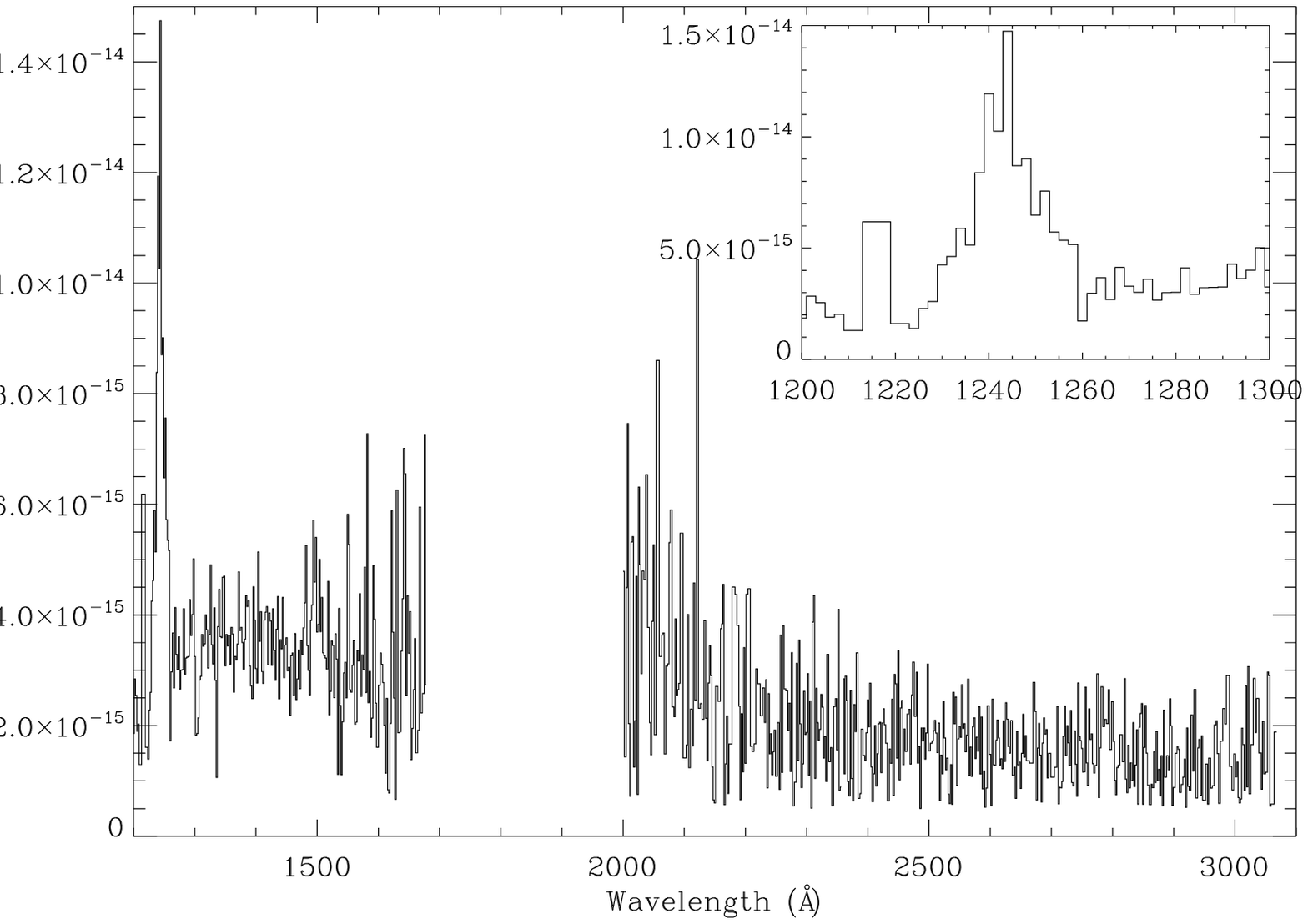}
\caption{The highest S/N portions of the UV spectrum from D42 rebinned 
to 2 \AA\ resolution.  The only prominent line is \ion{N}{5} at 1240 \AA\
which is shown in greater detail in the inset.
\label{stisv1}}
\end{figure}

Due to pointing constraints, the first visit could not be scheduled until
D42.  This observation used the STIS medium echelle grating to obtain
coverage from 1150 - 3100 \AA. Surprisingly, only one emission line,
\ion{N}{5} (1240\AA), was detected, see Figure \ref{stisv1}.  The continuum
was relatively flat with an integrated UV flux from 1140 - 3130 \AA\ of
2.8$\times$10$^{-12}$ erg cm $^{-2}$ s$^{-1}$ (uncorrected for extinction).

\begin{figure}
\plotone{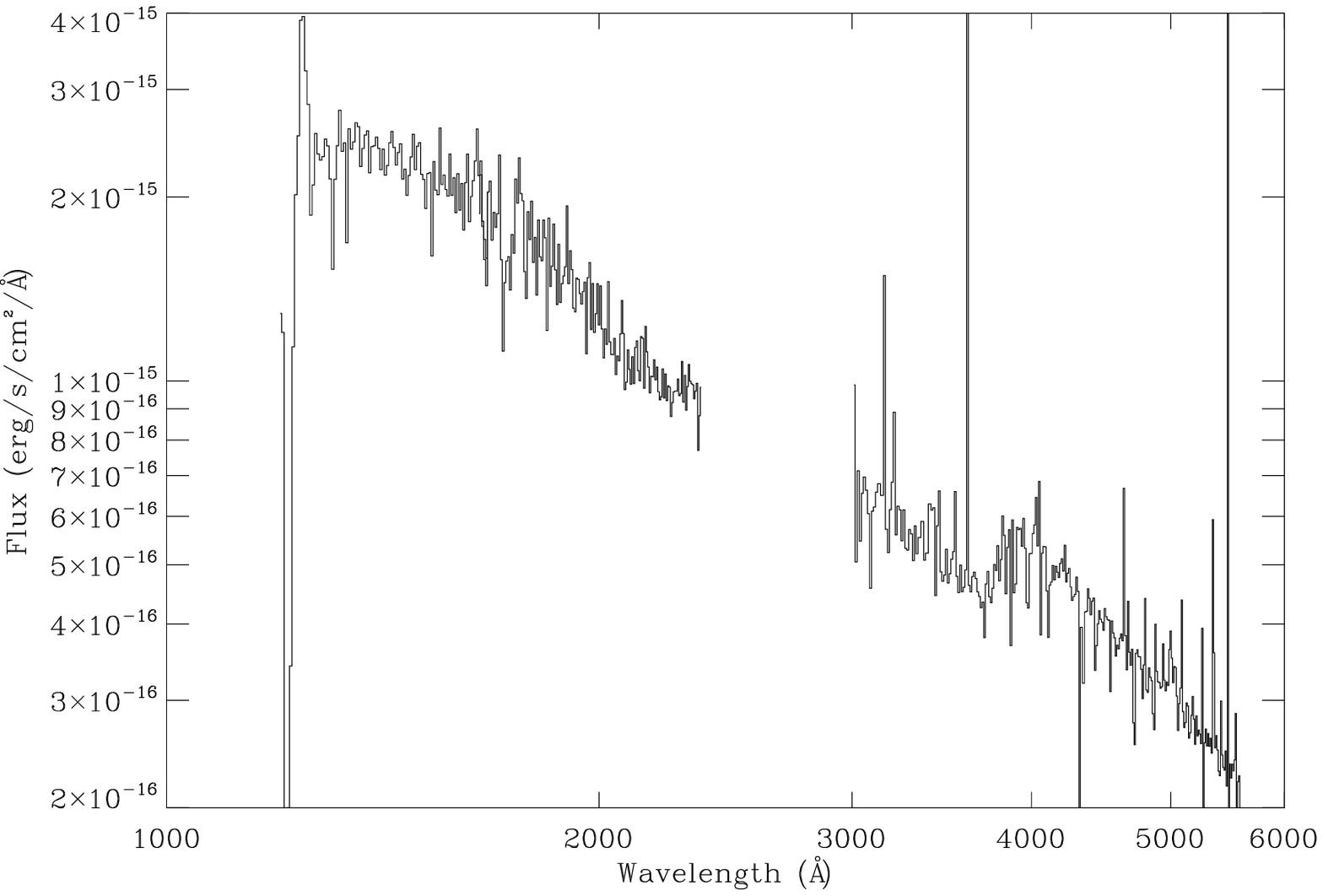}
\caption{STIS G140L, G230L, and G430L spectra binned to 5 \AA\ resolution 
from D57.  The only detected line is again \ion{N}{5} at 1240 \AA.
\label{stisv2}}
\end{figure}

For the second visit on D57 the low resolution grating was used since the
nova was already too faint to observe with the echelle.  The integrated
1140 - 3130 \AA\ UV flux had decreased to 1.8$\times$10$^{-12}$ erg cm
$^{-2}$ s$^{-1}$.  An optical grating exposure was also included since the
source could no longer be observed with the {\it SMARTS} spectrograph.  Except
for \ion{N}{5}, no emission lines were detected in the UV and optical
spectra.  In addition, the optical spectrum showed a Balmer discontinuity
that had not been present before which was likely due to contamination
from the field star.  Figure \ref{stisv2} shows the combined
D57 UV and optical spectra.

With the continuing decline in the light curve, the entire orbit allotment
was used for a single low resolution FUV exposure on D112.  Unfortunately,
neither the rapid decline nor the presence of the field star was 
anticipated prior to the observation, and {\it HST}'s acquisition locked on
the field star which was by then the brightest source in the field.

\subsection{SMARTS optical and near-IR data}

LMC 2012 was extensively observed spectroscopically and
photometrically with the {\it SMARTS} telescopes at Cerro Tololo \citep[see ][
for details]{Walt12}.  The spectroscopic observations were obtained between
D0.6 and D45.6.  LMC 2012 was photometrically monitored between D0.6
and D635. The cadence was initially daily but decreased as the source
faded.

\subsubsection{Photometry}

We obtained 250 photometric observations in BVRI/JHK from {\it SMARTS}. The
optical photometry is supplemented with 54 early time CCD BVRI observations
from the AAVSO. The optical and NIR photometry is also given in Table
\ref{phot}.

The mean optical rate of decay in the B and V bands was about 0.34 mag
d$^{-1}$ from D2 through D10 but about 0.04 mag d$^{-1}$ from D10
through D40.  The early decline rate was similar to that observed in the
UV uvm2 filter (Section \ref{uvot}) and the steady optical decay also
gave way to a variable and oscillatory behavior from D11 to D50.

After D80 the measured photometry was constant at about V=18.3 and B=18.2.
This is consistent with the BV photometry of \citet{Z04} for the field star
and indicates that LMC 2012 had faded below the optical brightness of the
field star after only three months.

There was no evidence in the optical or near-IR light curves for any dust
formation which is consistent with fast novae rarely forming extensive dust
shells \citep{G98}.

\subsubsection{Spectroscopy \label{smartsspectra}}

\begin{deluxetable}{crccc}
\tablecaption{{\it SMARTS} spectral observation log\label{optlog}}
\tablecolumns{7}
\tabletypesize{\scriptsize}
\tablewidth{0pt}
\tablehead{
\colhead{UT start time} &
\colhead{MJD start time} &
\colhead{$\Delta{t}$\tablenotemark{a}} &
\colhead{Exp.} & \colhead{Range} \\
\colhead{(YYYY-mm-ddThh:mm:ss)} & \colhead{(d)} &
\colhead{(d)} & \colhead{(s)} & \colhead{(\AA)}
}
\startdata
2012-03-26T23:55:54.4 & 56012.997 &  0.60 &  900 & 5620 - 6930 \\
2012-03-27T22:40:01.1 & 56013.944 &  1.55 &  900 & 3642 - 5412 \\
2012-03-28T23:37:34.7 & 56014.984 &  2.60 & 1200 & 5620 - 6930 \\
2012-03-30T23:28:13.6 & 56016.978 &  4.58 & 1200 & 5620 - 6930 \\
2012-04-01T23:53:07.8 & 56018.995 &  6.60 & 1200 & 3642 - 5412 \\
2012-04-02T23:30:22.9 & 56019.979 &  7.58 & 1200 & 3250 - 9400 \\
2012-04-05T20:20:10.5 & 56022.847 & 10.45 & 1200 & 3642 - 5412 \\
2012-04-06T23:38:40.7 & 56023.985 & 11.59 & 2700 & 5620 - 6930 \\
2012-04-07T23:34:19.3 & 56024.982 & 12.58 & 1200 & 3642 - 5412 \\
2012-04-08T23:20:47.0 & 56025.973 & 13.57 & 2700 & 3642 - 5412 \\
2012-04-12T20:37:01.3 & 56029.859 & 17.46 & 2700 & 5620 - 6930 \\
2012-04-14T23:21:57.2 & 56031.974 & 19.58 & 1500 & 3870 - 4540 \\
2012-04-15T20:55:30.7 & 56032.872 & 20.47 & 1800 & 3642 - 5412 \\
2012-04-16T23:32:48.5 & 56033.981 & 21.58 & 1800 & 5620 - 6930 \\
2012-04-18T23:27:37.3 & 56035.978 & 23.58 & 2700 & 3250 - 9400 \\
2012-04-24T21:20:00.7 & 56041.889 & 29.49 & 3600 & 3250 - 9400 \\
2012-05-10T23:25:35.7 & 56057.976 & 45.58 & 1800 & 3250 - 9400 \\
\enddata
\tablenotetext{a}{Where t$_0$ is the discovery date, 2012 March 26.397 UT (MJD 56012.897}
\end{deluxetable}

We obtained 17 optical spectra from D0.6 through D45.6, see Table
\ref{optlog}.  Unfortunately, LMC 2012 was too faint to observe with
the 1.5m telescope after conjunction with the Sun. 

The first (red) spectrum was obtained on D0.6. The H$\alpha$ line showed a
P-Cygni absorption profile due to the wind/expanding envelope at 
velocities ranging from -4500 to -5500 km s$^{-1}$ similar to the
description given in \citet{CBET3071P}.  No P-Cygni absorption components
were observed after D2.6.  The initial H$\alpha$ emission line showed a
FWZI of 247 \AA\ (11,300 km s$^{-1}$), an emission equivalent width of
308 \AA\, and integrated flux of 1.6$\times$10$^{-11}$ erg s$^{-1}$
cm$^{-2}$. Extremely broad lines, $\sim$ 5000 km s$^{-1}$, of \ion{N}{2}
(5755\AA) and \ion{He}{1} (5876\AA) were also present in the early red
spectra.

\begin{figure}
\plotone{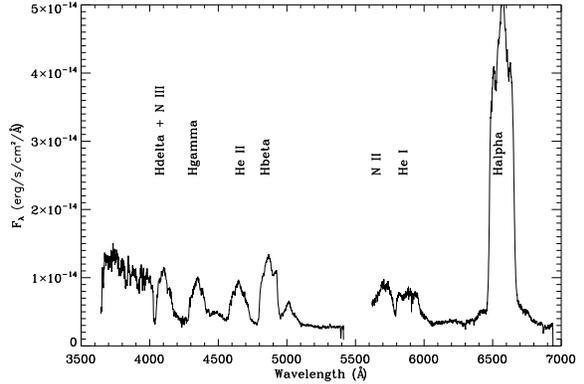}
\caption{Earliest combined blue and red {\it SMARTS} optical spectrum of 
LMC 2012 from the epoch D1.5 - D2.6.  The point-to-point uncertainties are of
order 1.5$\times$10$^{-15}$ erg s$^{-1}$ cm$^{-2}$ \AA$^{-1}$. 
Prominent lines are labeled.
\label{firstopt}}
\end{figure}

The first blue spectrum, D1.5, showed very bright emission at wavelengths
shorter than about 4100\AA, perhaps due to the confluence of the very broad
higher Balmer lines.  This, and the red spectrum obtained on D2.6,
is shown in Figure \ref{firstopt}.  The combined spectrum is similar to the
earliest spectra of the very fast ONe novae LMC 1990 \#1 \citep{Will91} and
V4160 Sgr \citep{Will94}.

To see how the expansion velocity in LMC 2012 compared to other novae, the
large, uniform sample of 52 Galactic and Magellanic Cloud novae with
measured FWHMs obtained near visual maximum from \citet{Sch11} was used.
\citet{CBET3071P} measured the FWHM of the H$\alpha$ line near maximum to
be 125 \AA\ (5700 km s$^{-1}$).  Only three novae, U Sco, V2478 Oph and
V2672 Oph, had greater FWHM at this time in the outburst. All
three are recurrent or suspected recurrent novae. Using the same criteria
in the LMC-only sample of \citet{Shafter13}, the FWHM of LMC 2012 is only
exceeded by two novae, LMC 1990 \#1 and LMC 1990 \#2. The former was a
very fast ONe type \citep{Van99} and the latter was a recurrent nova
\citep[LMC 1968;][]{Will91,Shore91}.

% ============================================================

By D2.6 the H$\alpha$ line had  the distinctive tri-partite line profile
common in U Sco-like recurrent novae. The central peak was the
strongest of the three peaks.  Four days later the blue spectrum was no
longer dominated by the Balmer lines, but by the Bowen \ion{N}{3} lines.
The strongest line in the low dispersion optical spectrum on D7.6 was
[\ion{Ne}{5}] (3426\AA), with the Bowen blend a close second.  \ion{He}{2}
(4686\AA) was not present on D6.6, but was strong on D10.4.  \ion{He}{2}
may have been present, but heavily blended, on D7.6. \ion{He}{2} was narrow
($\sim$ 25\AA\ or 1600 km s$^{-1}$), and was the strongest line in the blue
spectrum by D13.6.  A review of the optical spectra in \citet{Walt12} and
the X-ray light curves in \citet{Sch11} show that in fast novae, the narrow
\ion{He}{2} (4686\AA) emission appears before the emergence of soft X-rays.
The evolution of the narrow \ion{He}{2} line in LMC 2012 is consistent with
the SSS appearance prior to D18.

\begin{figure}
\plotone{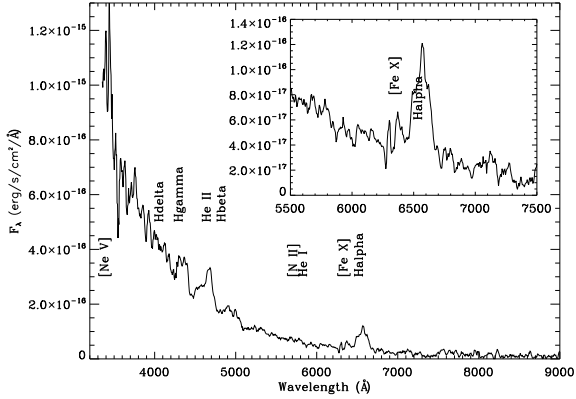}
\caption{Full {\it SMARTS} optical spectrum of D23.6, obtained
near the peak of the SSS emission.  Prominent lines are labeled.
\label{SSSopt}}
\end{figure}

By D11.6 there was some excess emission around 6400\AA\ that could be
associated with [\ion{Fe}{10}] (6375\AA).  This emission was present in the
red spectra until D29.5. If this excess were due to the actual
emergence of this emission line, it was consistent with the emergence of
the SSS on day D18 \citep{Sch11,K84}.  Figure \ref{SSSopt} shows the {\it
SMARTS} spectrum on D23.6.  By that time the nova had faded sufficiently
that only \ion{He}{2}, H$\alpha$, and possibly [\ion{Ne}{5}] and
[\ion{Fe}{10}] were still visible. In the last spectrum on D45.6 there were
no obvious emission lines.

\section{Modeling the SSS evolution}

\subsection{Period analysis \label{periodanal}}

\begin{figure}
\includegraphics[angle=-90,scale=0.30]{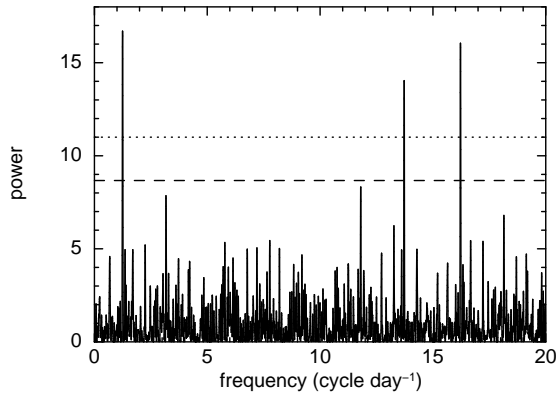}
\caption{Lomb-Scargle periodogram of the \Swift\ uvm2 photometry obtained
between D19 to D60 after subtraction of a first order polynomial, with
the 99.0\% and 99.9\% confidence levels shown. A significant peak is seen
at 1.2473 cycles day$^{-1}$ (=19.24 hours); other strong peaks are aliases of
this with the \Swift\ orbital period (15 cycles day$^{-1}$) and are not present
when the source periodic modulation is subtracted from the dataset.
\label{LS}}
\end{figure}

\begin{figure}
\includegraphics[angle=-90,scale=0.3]{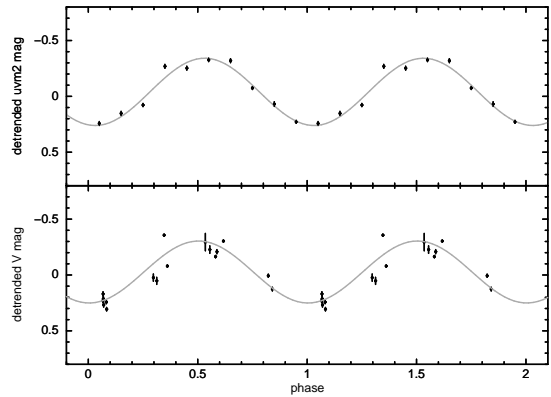}
\caption{(Top) The \Swift\ uvm2 photometry obtained between D19 to D60 after
subtraction of a first order polynomial folded at the 19.24 hour period;
errors represent the standard deviation of the distribution of values in
each 0.1 phase bin. (Bottom) The {\it SMARTS} V band photometry obtained between
D19 to D60 after subtraction of a first order polynomial folded at the
same period; errors are propagated. Solid curves represent the best sine
fits.
\label{uvotVphase}}
\end{figure}

A Lomb-Scargle periodogram (LSP) was formed from the 50 \Swift\ uvm2
photometric measurements obtained from D19 to D60 after subtraction of a
first order polynomial to remove the secular decline. A peak in the
periodogram at 1.2473 cycles day$^{-1}$ (Figure \ref{LS}), in excess of the
99.9\% confidence level of 11.0 and corresponding to a period P $= 19.24
\pm 0.03$~hours, is derived with the method of \citet{HB86} under the
reasonable assumption of even sampling. The error is derived from a least
squares sine fit to the de-trended data with the photometric errors
increased artificially to allow a fit with a reduced chi squared of unity.
The periodogram also shows aliases with the \Swift\ orbital period, as
expected from the convolution of the source signal with the window function
of the data, and a peak at 3.17 cycles day$^{-1}$ with a power
corresponding to $\sim$ 90\% confidence; these peaks are not present in a
periodogram of the dataset with the 19.24 hour modulation subtracted,
confirming that they are not intrinsic to the source. The uvm2 light curve
folded at the 19.24 hour period is shown in Figure \ref{uvotVphase};
modeled with a sine function, the amplitude is 0.306 $\pm$ 0.031
magnitudes. A periodogram of the D19-60 XRT 0.3-10 keV X-ray light curve
de-trended with a second order polynomial shows no significant power at the
UV period.  We find the 90\% upper limit to the amplitude of any modulation
of the X-rays at the UV period to be 15\%.  

The {\it SMARTS} light curve shows a modulation with the same period 
detected in the UV, although these data do
not permit the independent detection of this period. The 19.24 hour period
amplitudes in the BVRIJHK filters were 0.269 $\pm$ 0.007, 0.275 $\pm$ 0.009,
0.280 $\pm$ 0.011, 0.305 $\pm$ 0.019, 0.39 $\pm$ 0.31, 0.47 $\pm$ 0.26, and
0.76 $\pm$ 0.23 magnitudes, respectively over D19-60; 90\% confidence
errors are given. The {\it SMARTS} V band photometry folded at the 19.24 hour
period is also shown in Figure \ref{uvotVphase}.

The origin of the modulation is not known but it is likely orbital in
nature.  Even in a long period system, the secondary would be tidally
locked and strongly irradiated around the substellar point by the hot WD.
In an inclined system (see Section \ref{lineprofilefit}) the distended and
illuminated lobe would produce variations in the UV through NIR with
amplitudes similar to what was observed.  Conversely, the X-ray light curve
is constant (Sections \ref{xrtsection} and \ref{chandrasection}) as the
X-rays are emitted from the WD atmosphere which is not modulated by the 
orbital motion.

While it is possible that observed UV and optical variability could
also come from illumination and heating of a warped accretion disk, we
discount this possibility as it would not account for the lack of similar
modulation in the X-ray data sets.  In addition, an accretion disk would
have to either survive the initial nova explosion or reform very quickly
even as the WD was at peak luminosity and temperature. More exotic
scenarios may also be at work but an illuminated secondary is the simplest
explanation that fits all the available data and thus is favored in the
subsequent analysis.  Confirmation of a 19.24 hr orbit will require
observations during quiescence.

\subsection{Modeling the X-ray spectral evolution \label{swiftmodel}}

\btxt{We use the extensive \Swift\ XRT data set to model the entire X-ray
evolution during the outburst in LMC 2012.} Initially, the \btxt{\Swift} X-ray 
spectra were modeled using a combination of a
blackbody or a plane-parallel, static, non-local thermal equilibrium
atmosphere component \footnote{Grid \#011 from
\url{http://astro.uni-tuebingen.de/$\sim$rauch/TMAF/flux\_HHeCNONeMgSiS\_gen.html}
In the framework of the Virtual Observatory (VO;
\url{http://www.ivoa.net}), these spectral energy distributions are
available in VO compliant form via the VO service TheoSSA
(\url{http://vo.ari.uni-heidelberg.de/ssatr-0.01/TrSpectra.jsp?}) provided
by the German Astrophysical Virtual Observatory (GAVO;
\url{http://www.g-vo.org}).} \citep{Rauch03,Rauch10} to parameterize the
soft emission, plus a single temperature optically thin thermal plasma to
account for the emission at higher energies.  
\btxt{Although a gross simplification of the underlying physics, in low
resolution X-ray spectra such as the XRT, blackbody models are sometimes
used to characterize the temperature and luminosity changes of the soft
emission.  Blackbody fits, however, can underestimate the true temperature and
generally overestimate the bolometric luminosity \citep{1994A&A...288L..45H}.
A more realistic treatment of the physics comes from the use of hydrostatic
model atmospheres which can, unlike blackbodies, sucessfully fit the higher
resolution X-ray spectra. For LMC 2012, the}
C-stat values for the
blackbody fits were significantly worse than for the model atmosphere fits
and thus the blackbody fits were not used in the analysis.  In addition,
the choice of model atmosphere did not significantly affect the fit to the
data or the derived properties.  Both the hard and soft model components
were absorbed by a freely-varying column. Luminosities were calculated
assuming a distance of 48 kpc.

\begin{figure*}
\includegraphics[angle=-90,scale=0.6]{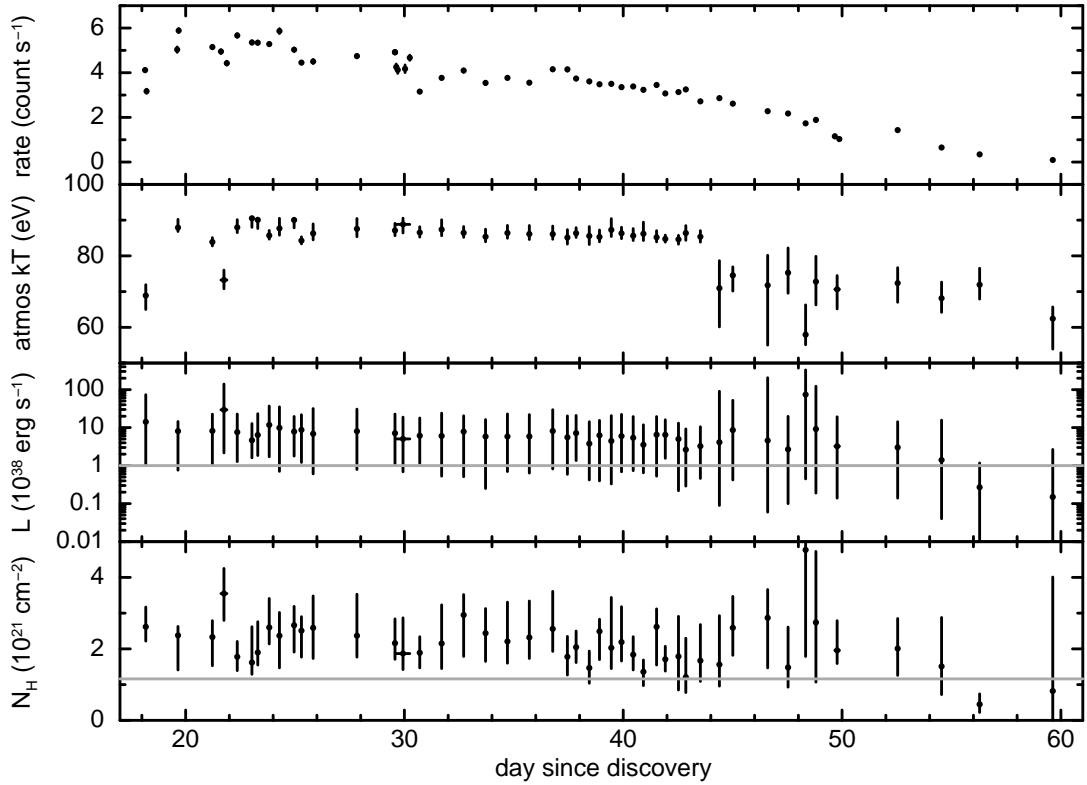}
\caption{Model atmosphere fits to the \Swift/XRT spectra during the time
LMC 2012 was in the SSS phase.  The top panel shows the XRT light curve
on a linear scale.  The next three panels give the best fit model atmosphere
effective temperatures, bolometric luminosities, and N$_H$ parameters.
The grey lines in the luminosity and N$_H$ panels represent a 
typical Eddington luminosity for a high mass WD (1$\times$10$^{38}$ erg
s$^{-1}$ and the extinction, 1.16$\times$~10$^{21}$~cm$^{-2}$, derived from 
the \Chandra\ model fit, respectively.
\label{swiftfit}}
\end{figure*}

Figure \ref{swiftfit} shows the results of the model atmosphere fits to the
\Swift\ X-ray data during the SSS phase. From D20 to D42 the X-ray spectra
are well fit by models with a constant effective temperature of order 86
eV, $\sim$ 1 MK.  The fitted model luminosities are not as well constrained
with values between (1-10)$\times$10$^{38}$ erg s$^{-1}$.  Since the
distance to the LMC is well known, an upper limit on the bolometric
luminosity can be established from the Eddington limit for a 1.4
M$_{\odot}$ WD, i.e. $\sim$ 1$\times$10$^{38}$ erg s$^{-1}$ cm$^{-2}$.
Excluding the models with the largest errors, the hydrogen column density
evolution is compatible with N$_H \sim$ 2$\times$10$^{21}$ cm$^{-2}$.  This
value is consistent with the external extinction along the line of
sight of N$_H$ = 0.7$\times$10$^{21}$ cm$^{-2}$ used to correct the field
star photometry and FUV spectrum \citep[E(B-V) = N$_H$/4.8$\times$10$^{21}$
and E(B-V) = 0.15 mag][]{Bohlin78}.

\begin{figure}
\includegraphics[angle=-90,scale=0.3]{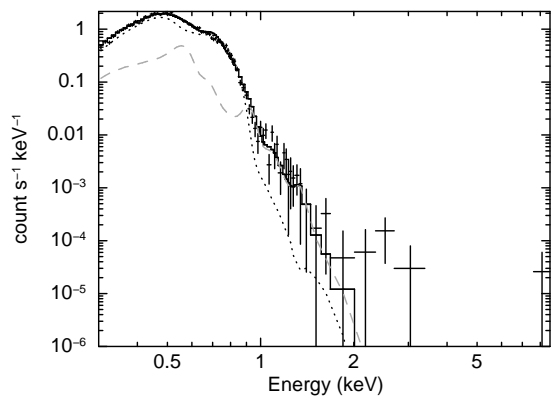}
\caption{The total XRT spectrum from the combined D18-D50 observations.
The best fit model atmosphere (dark dotted line) and MEKAL model (gray dashed 
line) are also shown.
\label{summedSSS}}
\end{figure}

The model sequence confirms that LMC 2012 was at its maximum effective
temperature early in the \Swift\ observations of the SSS phase and maintained
a constant bolometric luminosity at about the Eddington limit for
approximately 50 days.  Figure \ref{summedSSS} shows the combined XRT
spectrum from the D18-50 data.  The best fit had a model atmosphere
temperature of 86.2 $\pm$ 0.3 eV and an optically thin MEKAL component
temperature of 0.120 $\pm$ 0.007 keV.  The model N$_H$ was
1.7$\times$10$^{21}$ cm$^{-2}$ which is consistent with the typical LMC
N$_H$ value \citep{Welty12}.  

\begin{figure}[ht]
\includegraphics[angle=-90,scale=0.3]{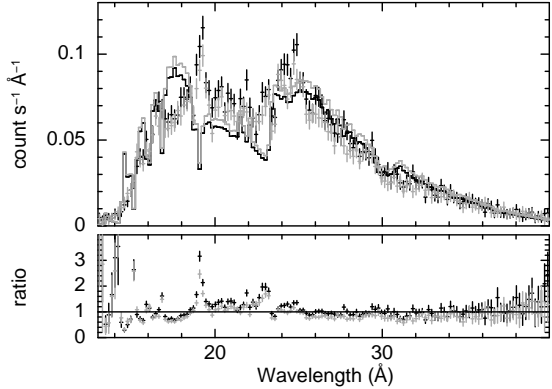}
\caption{The \Chandra/LETG spectrum (D32; black and gray plus symbols 
are the plus and minus orders, respectively) and the best fit Rauch atmosphere 
model from the same grid as used in the \Swift/XRT analysis.  The best fit 
model atmosphere temperature to the LETG spectrum, 88.0 $\pm$ 0.3 eV, was 
similar to that derived for the lower resolution XRT spectrum obtained on D31.7,
87$^{+3}_{-2}$ eV. Note that the significant residuals in the model fit
correspond to the strongest blue-shifted lines which are not incorporated
in the model.
\label{chandracomp}}
\end{figure}

\btxt{As an additional check on the validity of the models used to fit the
entire \Swift/XRT dataset}, 
the \Chandra/LETG spectra (plus and minus orders) were
fitted with the same atmosphere grid model as described above.  No
additional components were included. The resulting parameters for the
atmosphere component are consistent with those from the XRT data taken
close in time, although the best fit absorption column from the grating
spectra is lower, at (1.16 $\pm$ 0.03)~$\times$~10$^{21}$~cm$^{-2}$,
compared to 1.7$\times$10$^{21}$~cm$^{-2}$ from the combined \Swift\ D18-50
data spectrum or the (2.1$^{+1.1}_{-0.7}$)$\times$10$^{21}$~cm$^{-2}$ from
the D31.7 XRT spectrum, see Figure \ref{chandracomp}.  While
\citet{2012ApJ...756...43V} shows that at wavelengths longer than 40 \AA\
models are very sensitive to the choice of N$_H$, the signal-to-noise in the
\Chandra\ spectrum is not of sufficient quality in this region to constrain
N$_H$ further.

The lack of any significant hard X-ray emission in LMC 2012 is surprising
as most novae \btxt{bright enough for X-ray observations} have an early
period of hard X-ray emission from shocks \citep[e.g. see][for
details]{ATel2419,Sch11,2014ApJ...788..130C}. The shocks are thought to
arise from either internal shocks within the ejecta or the ejecta running
into pre-existing material such the wind from a red giant companion.
Fitting another MEKAL component centered at KT = 5 keV improves the fit to
the data above 1 keV in the combined spectrum shown in Figure
\ref{summedSSS}.  The 90\% confidence upper limit on the bolometric flux of
this new hard component is 8.2$\times$10$^{-14}$ (observed) or
1.3$\times$10$^{-13}$ (unabsorbed) erg cm$^{-2}$ s$^{-1}$.  This is
equivalent to bolometric luminosities of 9.8$\times$10$^{30}$ and
1.3$\times$10$^{31}$ erg s$^{-1}$, respectively, at 48 kpc.  This
luminosity upper limit is significantly lower than the 10$^{34-35}$ erg
s$^{-1}$ that is typically observed \citep{Mukai08,Metzger14}.

\subsection{Modeling the ejecta}
\subsubsection{Line profile fitting \label{lineprofilefit}}

\begin{figure}[ht]
\includegraphics[angle=0,scale=0.4]{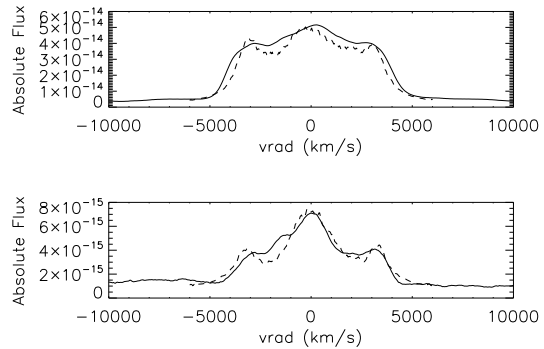}
\caption{Simulation (dotted lines) of two H$\alpha$ profiles from the 
{\it SMARTS} data (solid lines) using the Monte Carlo procedure from \citet{Shore13}.
The top spectrum is D2.6, the bottom is from D11.6.  The model parameters are 
$v_{max}$=6500 km s$^{-1}$, $\Delta R/R$=0.3, $i=55^{\arcdeg}$, 
$\theta_o=30^{\arcdeg}$, $\theta_i=70^{\arcdeg}$. 
See text for details. \label{profs}}
\end{figure}

To obtain an idea of the geometry of the ejecta, we modeled the optical
Balmer lines using the Monte Carlo procedure described in \citet{Shore13}.
Figure \ref{profs} shows two H$\alpha$ profiles compared with the model
parameters that were chosen to provide an approximate representation,
consistent with the dynamics and profile evolution.  A similar solution 
was obtained for the other Balmer lines.  The model parameters are the
relative shell thickness ($\Delta R/R$) where $R$ is the radius given by
the maximum observed velocity during the earliest stages, the inner and
outer angles of bipolar symmetric ejecta ($\theta_i, \theta_o$), and the
inclination of the axis of the ejecta to the line of sight $i$.  The
displayed profiles were smoothed to 100 km s$^{-1}$ to reduce the
stochastic fluctuations and the line was assumed to be formed by recombination.
We assumed a ballistic velocity law.  For $i < 50^o$ there is no
central peak and multiple low velocity peaks are obtained if $\theta_i <
70^o$.  Otherwise, with the ejecta appear to be bipolar
with a moderately high inclination to the line of sight, subtending a
solid angle of about 2$\pi$ with respect to a spherical shell.  No
spherical solution is acceptable at any time and the inner angle appears to
have decreased over time with increasing transparency.  The same behavior
has been found for other novae similarly modeled
\citep[e.g.][]{Ribeiro13a,Ribeiro13b,Shore12}.

\subsubsection{Photoionization analysis \label{cloudy}}

We used the \cldy\ \citep{Fer13} photoionization code to fit the
pan-chromatic data set for four separate dates, D7.5, D29.5, D42, and D57.
For a given set of input parameters, \cldy\ solves the equations of thermal
and statistical equilibrium and predicts both a continuum and emission line
ratios.  A \cldy\ model for a nova requires a set of input parameters for
the ejected shell and the photoionizing source.  The source parameters are
the luminosity and spectral energy distribution.  The shell parameters
are the geometry, structure, hydrogen density, and elemental
abundances relative to hydrogen.

The  \cldy\ models require a large number of parameters so it is desirable
to minimize the set either from the data or physical assumptions.  For the
ejecta, the inner and outer radii for LMC 2012 were assumed to be equal to
minimum and maximum ejection velocities of 1000 and 5000 km s$^{-1}$ times
the number of days since discovery.  The model filling and covering factors
were set to 0.1 and 1, respectively, which are typical for similar
photoionization analyses \citep[see ][ for examples]{Sch07}.  The radial
variation of the ejecta number density was assumed to be (r/r$_i$)$^{-3}$
so that the mass is constant in the shell  (ballistic expansion).  After
fixing the radii and ejecta structure, the only free shell parameter that
determines the ejecta mass is the hydrogen density at the inner radius,
r$_i$.  The lack of emission lines in the later spectra means that the
ejecta abundances could not be constrained and were left at their (default)
solar values for this analysis.  Similar models with a LMC abundance of
Z=0.33Z$_{\odot}$ were calculated but there was no appreciable difference
in the results. Therefore, our results are insensitive to the abundance
selection and, unfortunately, do not allow a determination of the ejecta
abundances or WD composition with the available data.

%\clearpage
\begin{figure*}[ht]
\begin{tabular}{cc}
\includegraphics[scale=0.45]{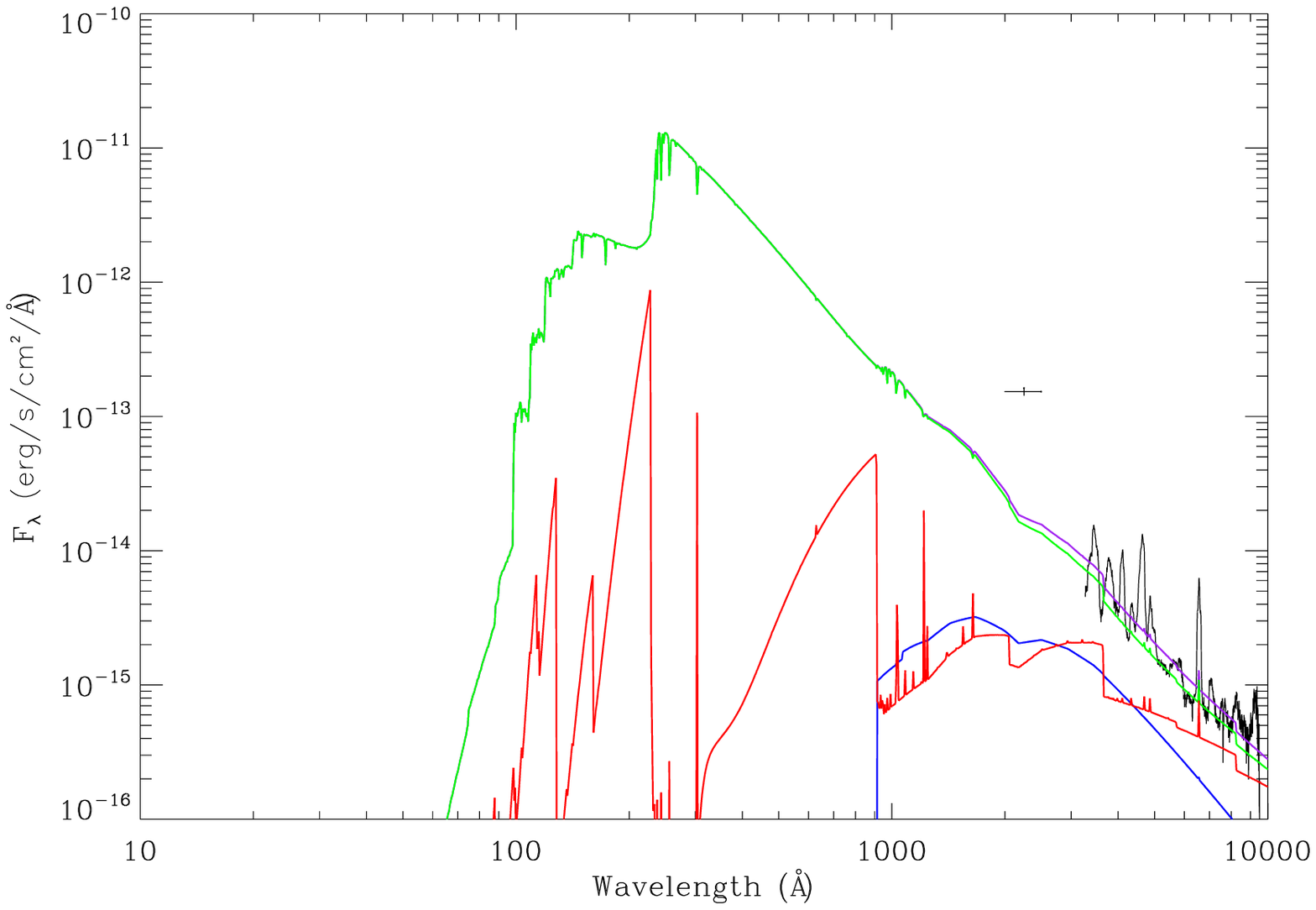} & \includegraphics[scale=0.45]{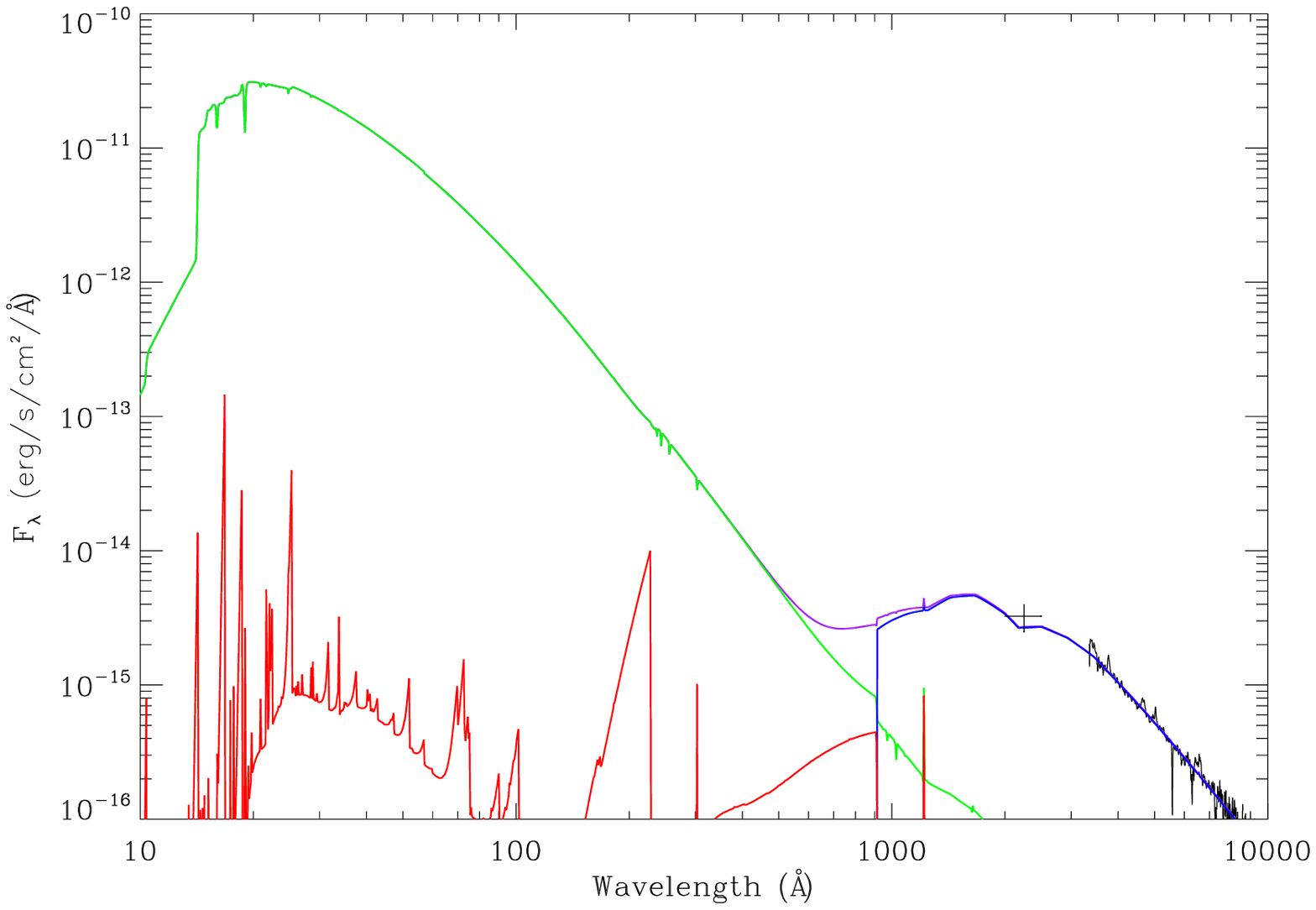} \\
\includegraphics[scale=0.45]{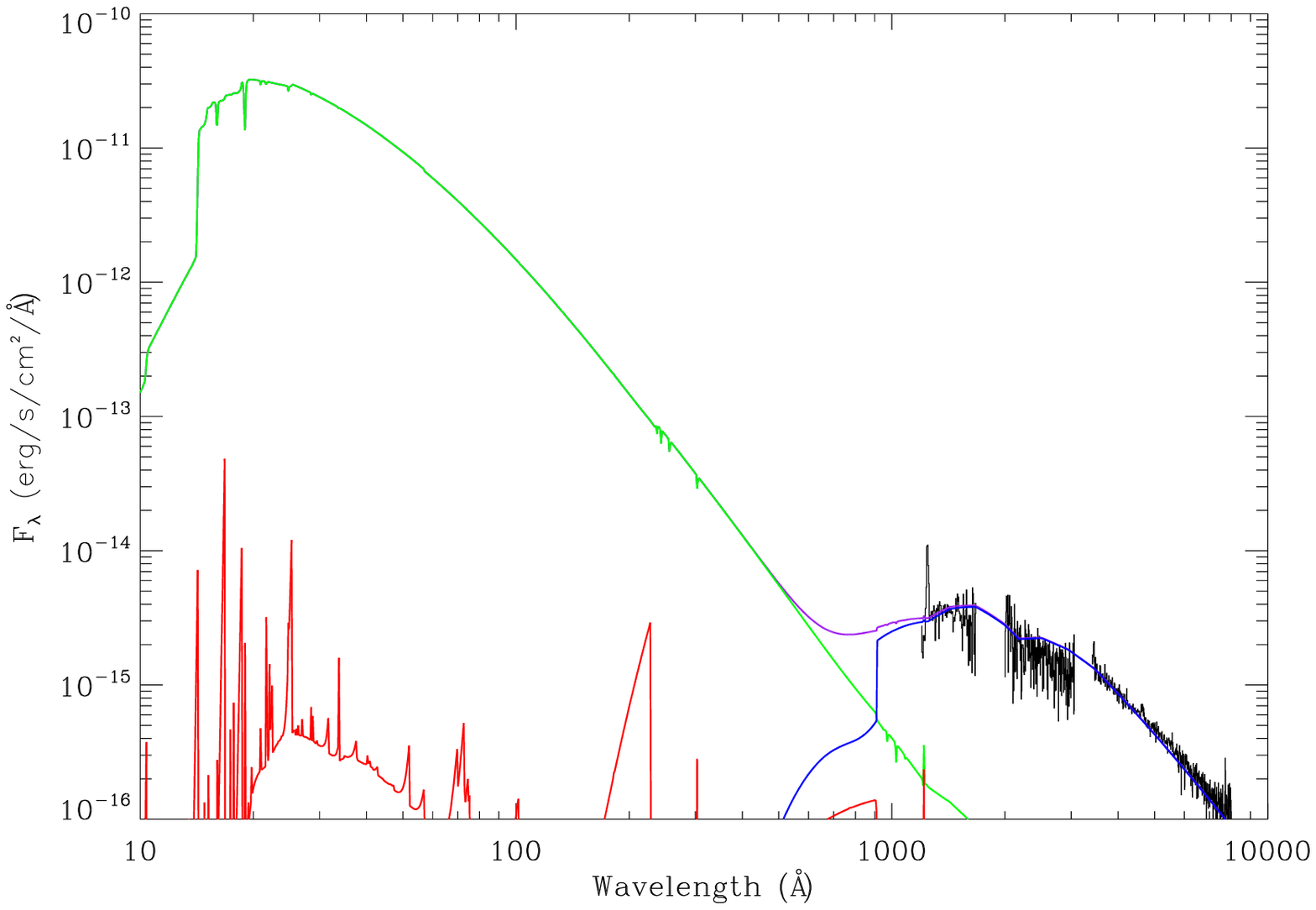} & \includegraphics[scale=0.45]{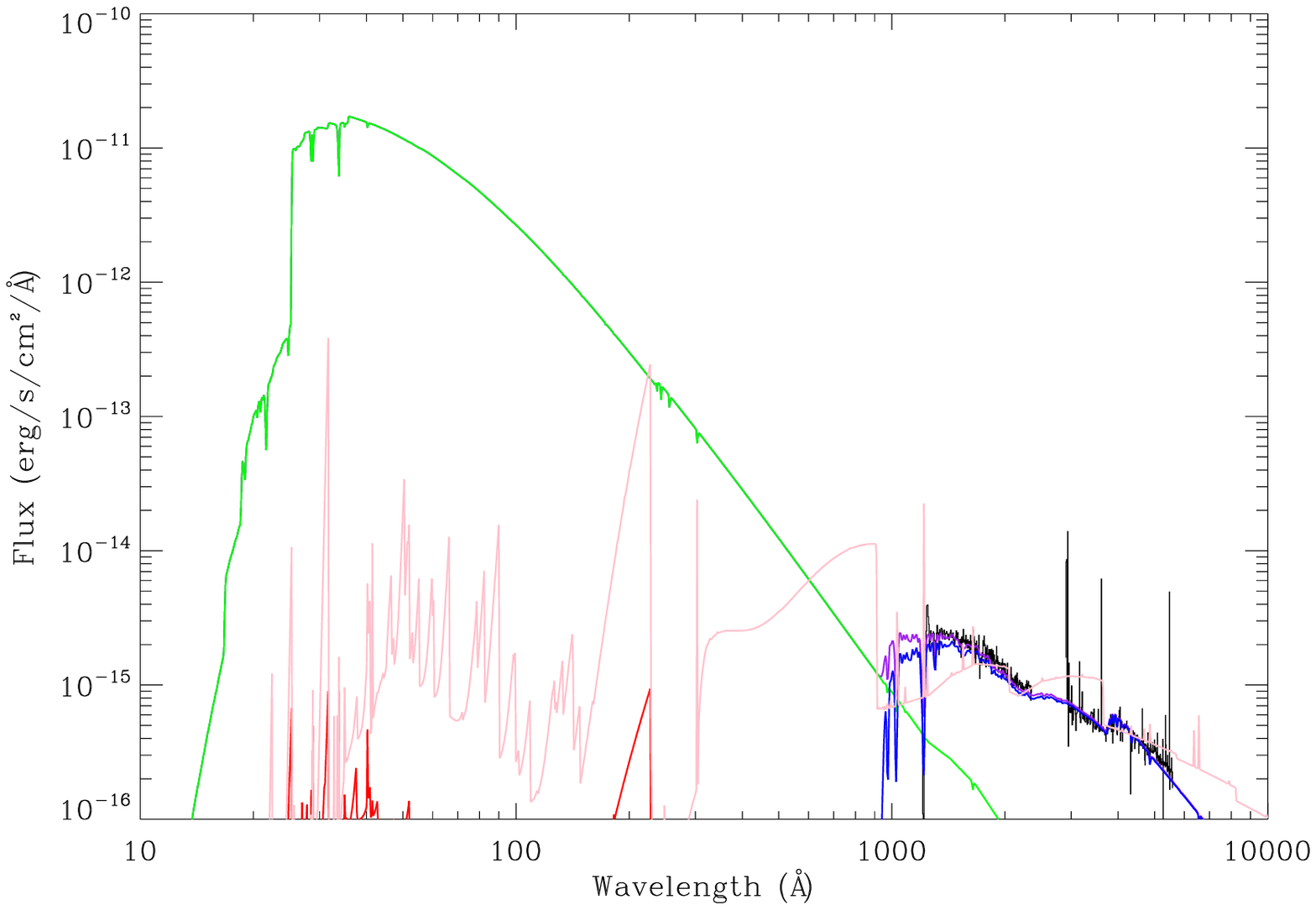} \\
\end{tabular}
\caption{The components of the best fit \cldy\ models to the D7.5 (a),
D29.5 (b), D42 (c), and D57 (d) rebinned UV and optical spectra, top left
to bottom right.  Similar to the estimated field star value, the \cldy\
models have been reddened by the mean LMC E(B-V) of 0.15 mag \citep{LMCEbmv}.
The model parameters are given in Table \ref{modelparms}. 
The green line shows the WD contribution, the blue line the
secondary source contribution, the red the ejecta contribution, and 
the purple the summed model. The pluses at 2246 \AA\ in the top two
figures show the flux from the uvm2 observation. The ejected shell component 
is only a significant contributor on the first date when the secondary is not
as luminous. The bottom right
figure also shows the contribution of a model with the same parameters 
except for 20X the ejecta mass in pink.  It is included to illustrate the 
amount of mass required to obtain sufficient UV/optical flux similar to the 
observations.  Note that the resulting model UV/optical continuum is a very 
poor fit to the observed spectra.
\label{modelWD}}
\end{figure*}

Photoionization of the ejecta results from the hot WD emission and thus the
temperature and luminosity are constrained by the modeling of the \Swift\ X-ray
data set.  As in Section \ref{swiftmodel}, Rauch model atmospheres were
used in the \cldy\ models.  Consistent with the \Swift\ and \Chandra\
results, a model with T$_{\rm eff}$ = 963,000 K and bolometric luminosity of
10$^{38}$ erg s$^{-1}$ was used for the D29.5 and D42 datasets.  Similarly,
a cooler and fainter model, with  T$_{\rm eff}$ = 638,000 K and L$_{\rm Bol}$ =
7.8$\times$10$^{37}$ erg s$^{-1}$, was used for the last modeled date, D57.
The predicted SEDs from the WD are shown in green in Figure
\ref{modelWD}.

We find that the best fit to the continuum from D7.5 which is also
consistent with the lack of a \Swift\ detection, used a Rauch model
atmosphere with T$_{eff}$ = 130,000 K and a slightly higher bolometric
luminosity of 2$\times$10$^{38}$ erg s$^{-1}$, see Figure \ref{modelWD}a.
The implied WD radius from this T$_{\rm eff}$ and L$_{\rm Bol}$ is about
0.1 of the orbital separation, see Section \ref{Dis}.  The D7.5 WD
parameters and N$_H$ = 5$\times$10$^{21}$ cm$^{-2}$ (for an ejected mass of
10$^{-6}$ M$_{\odot}$ 7 days after the outburst, see below) predict a
\Swift\ count rate of 10$^{-8}$ ct s$^{-1}$ in PIMMS which is consistent
with the observed upper limit of $<$0.008 ct s$^{-1}$ on D8.2. The uvm2
magnitude on the same day was converted to a flux \citep{Poole2008} and is
also shown in Figure \ref{modelWD}a. The WD model predicts less continuum
flux than is detected at the effective wavelength of the uvm2 filter.  The
likely explanation for this discrepancy is that the early NUV SED had
significant line emission, as seen in the D7.5 optical spectrum, that is
not reproduced by the WD model continuum.

At these model temperatures and luminosities, the WD contributes nothing to
the observed UV/optical SED except for the first date when the effective
temperature was much lower. Figure \ref{modelWD} shows the WD contribution
in green and the ejecta contribution in red.  Since the WD fits the
optical spectra during the first modeled epoch, a strong upper limit on the
model ejected mass can be established.  A model with an ejected mass
M$_{ej}$ = 1.4$\times$10$^{-6}$ M$_{\odot}$ provides the best fit to the
data and was adopted for the other three dates. This mass estimate is also
consistent with those derived in Section \ref{xrtsection}.

The \cldy\ fits for the latter three dates were extremely poor, independent
of realistic ejecta masses.  The predicted continuum was generally at least
100 times lower than the observed SED when using the mass derived
from the first modeled epoch.  Artificially raising the model ejected
mass to significantly larger values did increase the predicted
UV/optical continuum luminosity but the resulting recombination spectrum
was completely incompatible with the observed SED. Figure
\ref{modelWD}d shows an example the poor fit of a model where the mass was
increased by a factor of 20 to match the UV and optical spectra (pink line).

Since the contribution of the model ejecta could not fit the later
observations, another light source with peak flux in the UV was
required.  Building on the assumption that the UV/optical/NIR modulation
described in Section \ref{periodanal} was due to an illuminated secondary
star, we add this contribution to the model. Estimates of the effective
temperature of the secondary's ``day'' side can be made from geometric
arguments of the amount of flux intercepted \citep{Ext05} assuming that the
secondary is in thermal equilibrium.  \citet{Rap82} find that for
conservative mass transfer the mass ratio, q = M$_{sec}$/M$_{pri}$, is
$\leq$ 2/3.  To illustrate the expected illumination
temperatures under reasonable assumptions consistent with the data, we
adopt q = 2/3, a WD mass near the Chandrasekhar limit, and a 19.24
hour period to derive a secondary day-side temperature of 22,000 K.
This temperature in a blackbody or model atmosphere was used as the
starting point when fitting the UV and optical spectra in the last three
observations. The secondary SED contributions are shown in blue in Figure
\ref{modelWD}.

The best fits to the UV/optical data on D42 require a secondary blackbody
with this effective temperature and a luminosity of 5.9$\times$10$^{35}$
erg s$^{-1}$.  A black body was used due to the lack of a Balmer
discontinuity in the optical data.  In the D57 data the Balmer
discontinuity was present in the HST data and thus a cooler \citet{Kurucz}
model atmosphere with T$_{\rm eff}$ of 17,000 K and L$_{\rm Bol}$ =
2.5$\times$10$^{35}$ erg s$^{-1}$ was used.  Based on the uvm2 photometry
of Figure \ref{swiftm2lc}, the field star contamination during the second
{\it HST} visit was twice as large as in the first HST observation.  The
Balmer jump is likely from the field star as it contributed about 0.67
times the optical flux at this time.

The D42 blackbody temperature was used for the secondary in the D29.5 fit
but with a higher luminosity of 7.4$\times$10$^{35}$ erg s$^{-1}$. For the
D7.5 model, a temperature of 20,000 K and luminosity of
4.7$\times$10$^{35}$ erg s$^{-1}$ was used, which does not affect the fit
from the brighter WD primary, see Figure \ref{modelWD}a.

\begin{deluxetable}{lllll}
\tablecaption{\cldy\ model parameters\label{modelparms}}
\tabletypesize{\scriptsize}
\tablecolumns{2}
\tablewidth{0pt}
\tablehead{
\colhead{} & \multicolumn{4}{c}{Value} \\
\colhead{Parameter} & \colhead{D7.5} & \colhead{D29.5} & 
\colhead{D42} & \colhead{D57}
}
\startdata
WD T$_{\rm eff}$ & 130 kK & 963 kK & 963 kK & 638 kK \\
WD SED & Rauch log(g)=8 & Rauch log(g)=8 & Rauch log(g)=8 & Rauch log(g)=8 \\
WD L$_{\rm Bol}$ & 2.0$\times$10$^{38}$ erg s$^{-1}$ & 1.0$\times$10$^{38}$ erg s$^{-1}$ & 1.0$\times$10$^{38}$ erg s$^{-1}$ & 7.8$\times$10$^{37}$ erg s$^{-1}$ \\
2nd T$_{\rm eff}$ & 20 kK & 22 kK & 22 kK & 17 kK \\
2nd SED & Blackbody & Blackbody & Blackbody & ATLAS log(g)=4\tablenotemark{a} \\
2nd L$_{\rm Bol}$ & 4.7$\times$10$^{35}$ erg s$^{-1}$ & 7.4$\times$10$^{35}$ erg s$^{-1}$ & 5.9$\times$10$^{35}$ erg s$^{-1}$ & 2.5$\times$10$^{35}$ erg s$^{-1}$ \\
Initial H density & 3$\times$10$^9$ cm$^{-3}$ & 2.5$\times$10$^7$ cm$^{-3}$ & 2.0$\times$10$^7$ cm$^{-3}$ & 6.3$\times$10$^6$ cm$^{-3}$ \\
R$_i$ & 6.5$\times$10$^{13}$ cm & 2.5$\times$10$^{14}$ cm & 3.6$\times$10$^{14}$ cm & 5.2$\times$10$^{14}$ cm \\
R$_o$ & 3.2$\times$10$^{14}$ cm & 1.3$\times$10$^{15}$ cm & 1.8$\times$10$^{15}$ cm & 2.6$\times$10$^{15}$ cm \\
M$_{ejected}\tablenotemark{b}$ & 1.4$\times$10$^{-6}$ M$_{\odot}$ & 1.4$\times$10$^{-6}$ M$_{\odot}$ & 1.4$\times$10$^{-6}$ M$_{\odot}$ & 1.4$\times$10$^{-6}$ M$_{\odot}$ \\
\enddata 
\tablenotetext{a}{\citep{Kurucz} model atmosphere.}
\tablenotetext{b}{Upper limit.}
\tablecomments{The other \cldy\ parameters are a hydrogen density power law of
r$^{-3}$, filling factor of 0.3, covering factor of unity.
%and N and Ne enhanced by 100X and 10X, respectively.  
All abundances were kept at their solar abundances \citep{Asp05}.
}
\end{deluxetable}

The derived upper limit on the ejected mass is extremely small but is
consistent with the very early X-ray turn-on and turn-off times. The high
WD photoionization rate  with an extreme effective temperature and
bolometric luminosity on a very low mass shell produces highly ionized
ejecta.  The typical nebular lines were not observed in LMC 2012 because
ions such as \ion{O}{3}, \ion{C}{4}, \ion{N}{2}, and \ion{Fe}{7} are simply
not present.  The only emission lines that were observed during the nebular
phase, \ion{N}{5} and possibly [\ion{Ne}{5}] and [\ion{Fe}{10}], were from
high ionization potential species.  Without a large number of emission
lines, the \cldy\ models cannot constrain the ejecta abundances and thus
the composition of the WD could not be derived.  The final \cldy\ model
parameters are provided in Table \ref{modelparms}.

\section{Discussion \label{Dis}}

The available LMC 2012 data provide insight into the nature of the binary
system.  The rapid optical decline, early SSS detection, very short SSS
duration, bright SSS luminosity, high WD effective temperature,
large ejection velocities, and low estimated ejected mass are all 
consistent with a high mass WD, likely near the Chandrasekhar limit.

WD mass estimates can be found from various relationships established in
the literature.  The models of \citet{SA05} show that for a WD with
kT$_{\rm eff}^{max} \sim$ 86 keV and L = 1$\times$10$^5$ L$_{\odot}$ the
mass must equal or exceed 1.3 M$_{\odot}$ regardless of assumed WD plus
accretion composition mix.  Likewise, the \cite{Y05} models that best match
the observational parameters for LMC 2012 have a WD mass between 1.25 and
1.4 M$_{\odot}$ with an accretion rate about 1$\times$10$^{7-8}$
M$_{\odot}$ yr$^{-1}$.  \btxt{An additional mass estimate can be obtained
from the more recent WD modeling of \citet{Wolf13}.  For LMC 2012, the
turn-off time ($\sim$ 50 days) and maximum effective temperature imply a WD
mass between 1.30 and 1.34 M$_{\odot}$.  It should be noted that all of 
these models do not take into account all the parameters that are likely
to have a role amount of mass accreted and ejected so the WD mass derived
from the LMC 2012 values is only approximate.  Regardless, the 
available models are all consistent with a WD mass above 1.3 M$_{\odot}$.}

The observed modulation of 19.24 hours in the uvm2, BVRI and JHK
light curves from D20 to D60 is most likely associated with the orbital
period.  All the data presented in this analysis are best explained by the
illumination of the secondary day-side to effective temperatures of order
22,000 K by the hot WD primary in a relatively high inclination system.

Some limits on the inclination of the system can be established from the
available spectra.  The inclination has to be much less than 90$^{\arcdeg}$
since eclipses are not seen in either the X-ray or UV data.  The
inclination from the line profile analysis, 70$^{\arcdeg} > i >$
50$^{\arcdeg}$, is consistent with the absorption spectrum observed in the
high resolution \Chandra\ grating spectrum. \citet{Ness13} find that the type
of X-ray spectrum, absorption or emission, is determined by the system
geometry.  Emission line X-ray spectra are associated with high inclination
systems because the lines are from reprocessed emission in the accretion
disk.

Assuming q = 2/3, a Chandrasekhar mass WD, and a
19.24 hour orbital period, the system separation is 3.2$\times$10$^{11}$ cm
(4.8 R$_{\odot}$) and the secondary Roche lobe radius is
1.0$\times$10$^{11}$ cm (1.5 R$_{\odot}$).  In order to achieve mass
transfer, the secondary must have a radius equal to its Roche lobe radius
which implies a subgiant.  The inferred temperature from the UV and optical
continuum fit is much higher than expected for a late type subgiant.
However the day-side temperature in irradiated models can reach factors
of between four and ten times larger than the shadowed side 
\citep[e.g.][]{Waw09}.

A late type subgiant secondary at quiescence would be $>$ 4 magnitudes
fainter than the nearby field star and thus not observed in pre-outburst
surveys.  The outburst amplitude would be $>$ 10 magnitude which is also
consistent with the very fast t$_2$ time.

The mass accretion rate can be estimated assuming the mass loss rate of an
evolved secondary filling its Roche lobe and including magnetic stellar
winds \citep[see Eqn. 3.16 - 3.20 in][]{IF2008}.  The mass accretion rate
is 10$^{-8}$ M$_{\odot}$ yr$^{-1}$ using the same
assumptions as before, namely a Chandrasekhar mass WD, q = 2/3, and
P$_{orb}$ = 19.24 hr. Table 1 in \citet{G98} provides the WD envelope mass
necessary to reach the critical pressure to initiate a thermonuclear
runaway as a function of WD mass.  At the upper end, M$_{WD}$ = 1.35
M$_{\odot}$, the envelope mass is 4$\times$10$^{-6}$ M$_{\odot}$.  For the
estimated mass accretion rate, the time to obtain this envelope mass is
very short, 60 years.  This gives further credibility to the
hypothesis that LMC 2012 is a recurrent nova of the U Sco subclass.  It is
unlikely that archival searches of the LMC would turn up any previous
events as the 2012 outburst was only brighter than 16th magnitude in the V
band for about a week.

Maintaining a 10$^5$ L$_{\odot}$ bolometric luminosity for 50 days requires
1.4$\times$10$^{-7}$ M$_{\odot}$ of hydrogen to remain on the WD after the
initial explosion \citep{G98}.  This amount is 20\% of the estimated upper
limit on the ejected mass.  The ejected mass plus the WD mass burned is
still about four times less than the accreted mass for a 1.35 M$_{\odot}$
WD and suggests that the WD is growing in mass.  This makes LMC 2012 a
potential SN Ia progenitor \citep{Starr88,WS11,Starr14} assuming the WD is
not of the ONe class \citep[e.g.][]{Mason11}.  Unfortunately the lack of
many shell emission lines makes an abundance determination from the
available data problematic.

\section{Summary}

\begin{enumerate}

\item LMC 2012 had a very fast optical/UV decline and its optical spectral
evolution was similar to that of U Sco. The V band t$_2$ time of two days
was one of the fastest ever observed.  Detection of similar outbursts in
the LMC will require full time monitoring at a high cadence since this nova
was brighter than V = 16 mag for less than 8 days.  The telescopes and
detectors of most amateur astronomers are only sensitive to visual
magnitudes brighter than about 13 mag.

\item An expansion velocity of $\sim$ 5,000 km s$^{-1}$ was inferred from
P-Cygni absorption observed in the early optical spectra. Absorption
lines with similar blue shifts were also observed in the later \Chandra\ SSS
spectrum.

\item LMC 2012 evolved very rapidly in the X-ray band with turn-on and
turn-off times of $\sim$ 13 and 50 days, respectively.  Both X-ray
timescales are extremely short compared to most other Galactic
\citep{Sch11} and M31 novae \citep{Henze14}.  \btxt{To model the X-ray
evolution we fit all the \Swift\ observations with a series of
\citet{Rauch03} model atmospheres.  We confirmed this approach by
sucessfully fitting the single \Chandra\ observation with a model
atmosphere with similar parameters as determined from the \Swift\ data set
around the same time.}  The results reveal a very hot WD with 
\btxt{maximum effective temperatures of 86.2 $\pm$ 0.3 eV, $\sim$ 1 MK,}
during the SSS phase. This temperature is also one of the
highest ever found in a nova and similar to that of RS Oph \citep{Osb11}
and V745 Sco \citep{ATel5897}.  The X-ray luminosity from the model
atmosphere fits was constant from D20 to D50 at $\sim$ 1$\times$10$^{38}$
erg s$^{-1}$.

\item The UV, optical and NIR light curves all showed oscillatory behavior
during the X-ray SSS phase.  Using the \Swift\ uvm2 data, we find a period of
19.24 hours. The BVRI and JHK data sets can also be well fit with the same
period.  There is no similar periodicity in the X-ray light curve. 
The period derived from the UV-IR modulation is likely orbital in nature.
The line profile fitting of the H$\alpha$ line provides an inclination 
estimate of 60$\pm$10$^{\arcdeg}$ which is consistent with both the modulation
amplitudes observed in the various filters and the strong absorption lines
detected in the \Chandra\ grating observation \citep{Ness13}.

\item An extremely unusual discovery was that the UV spectra only showed
the \ion{N}{5} line at 1240\AA.  Even in the optical, the emission lines
quickly faded as the nova progressed to the SSS phase showing no nebular
lines and weak, if any, coronal lines.  The puzzling lack of lines can be
explained by the high ionization of the low mass ejecta which was largely
ionized by a hot and luminous WD.  The very low upper
limit on the hard X-ray, $>$ 1 keV, luminosity is also consistent with a
small ejection mass since there is less material to be involved in shock
emission.

\item All the observed UV and optical continuum data can be fit with a
binary system model consisting of a hot WD, \btxt{whose photoionization
parameters are derived from the fits to the X-ray data set,} ionizing a
very small amount of ejected material ($\sim$ 1$\times$10$^{-6}$
M$_{\odot}$) and illuminating a secondary source.  The extremely small
derived ejecta mass contributes essentially nothing to the later observed
spectral energy distribution. The contribution from the secondary is
primarily responsible for the later UV/optical SEDs whereas the earliest
optical spectra are consistent with the Rayleigh-Jeans tail of a WD
photosphere too cool at that time to be detectable in X-rays.

\item The rapid X-ray, UV, and optical evolution, the large expansion
velocities seen throughout the outburst, plus the low mass ejected 
imply LMC 2012 is a recurrent nova of the U Sco subclass occurring on a
high mass WD in a moderately long period system with a high mass accretion
rate.  The available evidence implies that the WD is gaining mass
every outburst.  Unfortunately, the lack of significant line emission in
the UV and optical spectra did not allow us to determine the ejecta
abundances and thus the WD composition could not be inferred.  Future
modeling of the \Chandra\ spectrum may provide the necessary insights on
the WD composition.

\end{enumerate}

\acknowledgments

This research has made use of data obtained from NASA's \Swift\ satellite.
We thank Neil Gehrels and the \Swift\ team for generous allotments of ToO
and fill in time.  We thank Harvey Tananbaum for the \Chandra\ Director's
Discretionary Time observation of  LMC 2012.  Support for HST Program
number 12484 was provided by NASA through a grant from the Space Telescope
Science Institute, which is operated by the Association of Universities for
Research in Astronomy, Incorporated, under NASA contract NAS5-26555.  Stony
Brook University's initial participation in the {\it SMARTS} consortium was
made possible by generous contributions from the Provost of Stony Brook
University.  This research has been supported in part at the University of
Chicago by the National Science Foundation under Grant PHY 08-22648 for the
Physics Frontier Center "Joint Institute for Nuclear Astrophysics" (JINA).
We acknowledge with thanks the variable star observations from the AAVSO
International Database contributed by observers worldwide and used in this
research.  KLP, JPO \& APB acknowledge the support of the UK Space Agency.
SS acknowledges partial support from NASA and NSF grants to ASU.  Finally,
we thank the referee for a thoughtful report that greatly improved the
paper.

{\it Facilities:} \facility{HST(STIS)}, \facility{Swift(XRT and UVOT)}, 
\facility{AAVSO}, \facility{CTIO:1.3m}, \facility{CTIO:1.5m}, \facility{CXO}

\end{document}